\documentclass[12pt,aps,amsmath,latexsym,amsfonts]{JHEP3}

\usepackage{bm}
\usepackage{graphicx}
\usepackage{amssymb}
\usepackage{amsmath}
\usepackage{cancel}

\usepackage{epsfig}

\newcommand{\be}{\begin{equation}}
\newcommand{\ee}{\end{equation}}
\newcommand{\bea}{\begin{eqnarray}}
\newcommand{\eea}{\end{eqnarray}}

\def\half{\textstyle{\frac{1}{2}}}

\def\H{\mathcal{H}}

\def\m{{\rm m}}

\title{Lifshitz black holes in IIA supergravity}
\author{
Luke Barclay$^1$, 
Ruth Gregory$^{1,2}$\thanks{Email: r.a.w.gregory@durham.ac.uk} , 
Susha Parameswaran$^3$\thanks{Email: susha.parameswaran@itp.uni-hannover.de} , 
Gianmassimo Tasinato$^4$\thanks{Email: gianmassimo.tasinato@port.ac.uk} ,
Ivonne Zavala$^5$\thanks{Email: e.i.zavala@rug.nl}\\
$^1${\it Centre for Particle Theory, South Road, Durham, DH1 3LE, UK}\\
$^2${\it Perimeter Institute, 31 Caroline St, Waterloo, Ontario N2L 2Y5, 
Canada}\\
$^3$\it{Institute for Theoretical Physics, Leibniz University Hannover,
Welfengarten 1, 30167 Hannover, Germany}\\
$^4$\it{Institute of Cosmology \& Gravitation, University of Portsmouth,  
PO1 3FX, UK}\\
$^5$\it{Centre for Theoretical Physics,
University of Groningen,
Nijenborgh 4, 9747 AG Groningen, The Netherlands }
}

\abstract{
We compute string theoretic black hole solutions having Lifshitz 
asymptotics with a general dynamical exponent $z>1$.  We start 
by constructing solutions in a flux compactification of six 
dimensional supergravity, then uplift them to massive 
type IIA supergravity.  Alongside the Lifshitz black holes we 
study the simpler anti-de Sitter solutions, of which there are
a 1-parameter family in this supergravity, and compare and 
contrast their properties.  The black holes are characterized by 
a two-form and scalar charge, and we numerically explore their 
configuration space and thermodynamical aspects. 
}

\preprint{DCPT-12/13}

\begin{document}
\newcommand{\zed}{$\mathbb{Z}_2$}

\section{Introduction}

The use of holographic methods to explore strong coupling in gauge theories
has yielded a particularly fruitful interaction between string theory
and low temperature physics. The typical set-up uses the concept of
gauge/gravity duality \cite{Malda} in which a classical gravitational
system with negative spacetime curvature has, on its boundary, equivalent
degrees of freedom to a strongly coupled gauge theory. Temperature
is gravitationally introduced into these systems by adding a black hole
in the bulk spacetime, and different holographic dual theories can
be constructed by having additional bulk fields (see \cite{CMP} for
reviews of this approach).

The standard application of gauge/gravity duality is the adS/CFT 
correspondence, which yields a generally scale invariant boundary theory,
however, more recently, attention has focussed on systems having
more general scaling properties, such as non-relativistic field
theories, \cite{SCH}, or, pertinent to this investigation, a general
dynamical Lifshitz scaling, $z$:
\be
t\to\lambda^z t\;\;\;, \;\;\;\; 
x^i\to\lambda x^i\;\;\;, \;\;\;\; r\to r/\lambda.
\label{lifscal}
\ee
In order to produce such a dynamical scaling, the spacetime metric 
must be posited to have the following form
\be
ds^2=L^2\left(r^{2z}dt^2-\frac{dr^2}{r^2}-r^2{dx_i
dx^i}\right),\label{eq:LifshitzMetric}
\ee
which explicitly respects the scaling (\ref{lifscal}).
In this metric, not only the asymptotics, but the full spacetime has
the required scaling symmetry. Clearly, such a spacetime requires a
matter content to produce this asymmetry, and this was first set out
in the paper of Kachru et al.\ \cite{KLM}, in which charges and 
fluxes of topologically coupled gauge fields provided the necessary
scaling. This theory is in fact on-shell equivalent to a somewhat simpler 
massive vector theory \cite{Taylor}, although the $r\to 0$ singularity
of these spacetimes exhibits certain pathologies \cite{CM}. 

As with any holographic theory, although we can explore empirical 
simple models, in order to have confidence that there is indeed a 
holographically dual field theory we should be able to construct a
qualitatively similar ``top down'' theory with Lifshitz scaling within
string theory. After initial halted progress, string theory embeddings 
of Lifshitz geometries with dynamical exponent $z=2$ were found 
in \cite{STLIFe,STLIFl}, by 
making a consistent massive truncation 
of type IIB supergravity to a lower dimensional theory resembling 
the phenomenological construction of \cite{KLM}.  Soon after, a 
method for constructing Lifshitz spacetimes within string theory for
arbitrary scaling exponent $z>1$ was put forward in \cite{GPTZ}. In
this approach, the Lifshitz space is constructed from 
a simple flux compactification
of Romans' gauged supergravity in five and six dimensions, 
\cite{romans6d,romans5d}, generalising the classic adS compactifications 
of those theories.  The lower dimensional supergravity theories 
can be obtained by dimensionally reducing type IIA or IIB 
supergravity, as shown in \cite{IIAred,IIBred}, and any solutions 
can immediately be uplifted to ten dimensions\footnote{See \cite{Singh} for other examples of non-relativistic solutions in massive type IIA supergravity.}. Further  Lifshitz and AdS
solutions in gauged supergravity and string theory have been also studied in \cite{HPZ}.

In order to explore physical dualities, we need to be able to set our
system at finite temperature, in other words, we need to introduce a 
black hole to our spacetime. 
Black holes in asymptotically Lifshitz spacetimes were initially hard to
build.  
However by now several such solutions have been found in 
simple phenomenological models, starting with the numerical work of
\cite{Danielsson:2009gi}.  By engineering a matter or gravity content to 
source the desired geometry, some analytical solutions have also been 
constructed.  Overall, Lifshitz black hole solutions in phenomenological 
models include numerical and analytic
studies, fixed as well as arbitrary critical exponents, 
horizons with various topologies, extensions to
other dimensions, higher-order theories of gravity, and Brans-Dicke
models \cite{Pang,Vandoren, LiBHpheno,R^2}.

As to embedding Lifshitz black holes into string theory, this 
can now be done following on from the string constructions of 
pure Lifshitz geometries discussed above.
Recently, numerical string Lifshitz black holes with dynamical 
exponent $z=2$ were presented in  \cite{Amado:2011nd}.  As 
in \cite{STLIFe,STLIFl}, their method was to identify a consistent 
massive truncation from type IIB supergravity to a lower dimensional 
model resembling previous phenomenological constructions.  In the 
present paper, we construct string Lifshitz black holes 
with general dynamical exponent $z>1$, generalising the lower 
dimensional supergravity/type IIA Lifshitz solutions, which 
were found in \cite{GPTZ} by deforming adS solutions.  
Alongside the asymptotically Lifshitz black holes, we study  
related asymptotically adS black holes;  thus we are able 
to draw on the intuition gleaned from the latter as well as 
identify which properties belong uniquely to the Lifshitz case. 

With holographic condensed matter applications in mind, our 
interest is in planar black hole geometries, whose boundary 
field theory propagates in flat 2+1 spacetime, and moreover 
we consider static geometries corresponding to equilibrium phases.
Naturally, the black hole solutions are not so simple as 
their pure adS or Lifshitz cousins, with the exception of 
adS-Schwarzschild.  By exciting the supergravity fields 
about this latter background in the probe limit, we learn about 
the charges inherent to our system.  Further progress can 
be made by expressing the supergravity field equations as 
an autonomous  dynamical system, whose fixed points are 
pure adS or Lifshitz.  By perturbing close to the fixed points 
we can understand in detail how general interiors, including 
black holes, can flow to the adS/Lifshitz asymptotics. Indeed, 
this method allows us to 
analytically characterize all the possible asymptotic behaviours
for static adS and Lifshitz black holes for our theory and
moreover helps to numerically integrate to the full 
black hole solutions.  

As might be expected, since our 
string/supergravity setup contains more degrees of freedom 
than the simple phenomenological models, the black holes 
have a rich structure.  In the end,  the black holes we find 
necessarily have some non-trivial scalar field, and aside 
from the horizon size, are characterized by two parameters, which can 
be interpreted as a form field charge and scalar charge.  
Thus we can begin to explore their configuration space, how 
the field profiles and thermodynamical properties change as the 
charges and dynamical exponent $z$ vary.

The paper is organised as follows.  In the next section, we 
present the six dimensional supergravity theory where our black holes 
will be found, its pure adS and Lifshitz solutions and the general 
planar geometries that we will study.  In section 3 we proceed to 
analyse this setup in detail, starting with some approximate 
analytic solutions, both through the whole spacetime but close 
to the adS-Schwarzchild black hole in the probe limit, and in 
the far field limit close to the asymptotic adS and Lifshitz 
geometry.  We then build upon these results in section 4, to 
find numerical solutions describing adS and Lifshitz black holes, 
uplift them to type IIA supergravity, and study their behaviour.  
In section 5 we briefly discuss thermodynamical properties 
of the black holes and  we conclude in section 6.  
We give the details of the dynamical 
system we use to solve the supergravity field equations in 
appendix A.  Finally, as an aside, in appendix B we identify 
some exact analytical black hole solutions in a generic 
dilatonic model, which could plausibly be related to a supergravity theory.

\section{The System}\label{sec:model}

In this section, we introduce the supergravity theory that 
will be the subject of the paper, present its pure adS 
and Lifshitz solutions, and propose the general Ansatz which we 
 use to find planar black holes that asymptote adS and Lifshitz geometries.  
We consider six dimensional $\mathcal{N}=4$ gauged supergravity, 
first presented by Romans
in \cite{romans6d}.  This theory can be obtained from a consistent 
truncation of massive Type IIA supergravity, and thus the solutions of 
the six dimensional theory can be uplifted to solutions 
in string theory \cite{IIAred}.  
In \cite{GPTZ} it was found that 
the field content and couplings of this theory  admit Lifshitz solutions.

The bosonic field content 
of 6D Romans' supergravity consists of the metric, $g_{AB}$,
a dilaton, $\phi$, an anti-symmetric two-form 
gauge field, $B_{AB}$, and a set of gauge vectors, 
$(A_A^{(i)},{\mathcal A}_A)$ for the gauge 
group $SU(2)\times U(1)$.  The bosonic part of the 
action for this theory is
\bea
S &=&\int d^6 x  \sqrt{g_6} \Biggl [
 -\frac14 R_6 +\frac12 (\partial \phi )^2
-\frac{e^{-\sqrt{2} \phi}}{4}\,
\left( {\cal H}^2 +F^{(i)2}\right)
+\frac{e^{2\sqrt{2}\phi}}{12} G^2 \nonumber \\
&&-\frac18 \, \varepsilon^{ABCDEF}\,
B_{AB} \left( {\cal F}_{CD} {\cal F}_{EF}
+{\rm m} B_{CD} {\cal F}_{EF}
+\frac{{\rm m}^2}{3} B_{CD} B_{EF}
+F^{(i)}_{CD} F^{(i)}_{EF} \right)\nonumber\\
&&+\frac18\left( g^2 e^{\sqrt{2} \phi}
+4 g {\rm m} e^{-\sqrt{2} \phi}-{\rm m}^2
 e^{-3 \sqrt{2} \phi}
\label{6dL}
\right) \Biggr ] \,, \label{act6ds}
\eea
where $g$ is the gauge coupling, m is the mass of the 2-form field $B_{AB}$,
${\cal F}_{AB}$ is a U(1) gauge field strength, $F^{(i)}_{AB}$ a nonabelian
SU(2) gauge field strength, and ${\cal H}_{AB} 
= {\cal F}_{AB} + {\rm m} B_{AB}$. 
Spacetime indices $A,B,...$ run from $0$ 
to $5$, and $\varepsilon$ is the Levi-Civita tensor density.  
Notice the presence of Chern-Simons terms in the previous action, 
identified in \cite{KLM} as an important ingredient
for the existence  Lifshitz configurations. 

Varying the action gives the equations of motion:
\be
\begin{aligned}
R_{AB} &= 2 \partial_A \phi \partial_B \phi
+ \frac12 g_{AB} V(\phi) +
e^{2\sqrt{2}\phi} \left(G_A^{\,\,\,CD}
G_{BCD} - \frac16 g_{AB}
G^2 \right) \\
& - e^{-\sqrt{2}\phi} \left(2 \H_A^{\,\,C} \H_{BC}
+ 2 F_A^{iC} F_{BC}^i
- \frac14 g_{AB} \left(\H^2
+ (F^i)^2 \right) \right) 
\end{aligned}
\label{6deom1}
\ee
\bea
\Box \phi &=& \frac12 \frac{\partial V}{\partial\phi} + \frac13
\sqrt{\frac{1}{2}} e^{2\sqrt{2}\phi} G^2 + \frac12
\sqrt{\frac12} e^{-{\sqrt{2}\phi}} \left(\H^2
+ (F^{(i)})^2\right) \\
\nabla_B \left( e^{-\sqrt{2}\phi} \H^{BA} \right)
&=& \frac16 \, \epsilon^{ABCDEF} \H_{BC} G_{DEF} \\
\nabla_B \left( e^{-\sqrt{2}\phi} F^{(i)BA} \right)
&=& \frac16 \, \epsilon^{ABCDEF} F^{(i)}_{BC} G_{DEF} \\
\nabla_C \left(e^{2\sqrt{2}\phi} G^{CAB} \right)
&=& - {\rm m} e^{-\sqrt{2}\phi} \H^{AB} - \frac14 \,
\epsilon^{ABCDEF} \left(\H_{CD} \H_{EF} 
+ F_{CD}^{(i)} F_{EF}^{(i)} \right) \,,
\label{6deom5}
\eea
where we have defined the scalar potential function:
\be
V(\phi) = \frac{1}{4} \left ( g^2 e^{\sqrt{2}\phi} 
+ 4 {\rm m} g e^{-\sqrt{2}\phi}
-{\rm m}^2 e^{-3\sqrt{2}\phi} \right ) \,.
\ee

Analogously to 
the Romans' solution adS$_4 \times H_2$ \cite{romans6d}, it was 
shown in \cite{GPTZ} that one can have a Li$_4\times H_2$
dimensional reduction of this 6D supergravity to a 4D Lifshitz space with
an internal hyperbolic manifold threaded by non-abelian magnetic flux.
The solution is given by
\be
ds^2=L^2\left(r^{2z}dt^2-r^2dx_1^2-r^2dx_2^2-\frac{dr^2}{r^2}  \right)
-\frac{a^2}{y_2^2}(dy_1^2+dy_2^2), \label{eq:6dMetric}
\ee
where $a$ is a constant, the radius of curvature of the hyperboloid which can 
be taken to be compact (see \cite{MN} for details). The
dilaton is also chosen to be  constant, $\phi = \phi_0$,
and the field configurations are
\be
\begin{aligned}
F^{(3)}_{tr}\,&=\, q \, b L^3 e^{\sqrt{2}\phi_0} r^{z-1}& 
& & &F^{(3)}_{y_1y_2}= \frac{q} {y_2^2} \\
G_{x_1 x_2 r}&= b L^3 r  &\Rightarrow& & & B_{x_1 x_2}\,
=\,\frac{b}{2} L^3 r^2  \, .
\end{aligned}
\label{eq:FieldStrength1}
\ee
The relations between the various constants are somewhat 
simplified by performing  the following rescalings
\be
\begin{aligned}
&\hat b = L b e^{\sqrt{2}\phi_0} \qquad \hat q = L e^{-\phi_0/\sqrt{2}} q/a^2\\
\hat g = L g e^{\phi_0/\sqrt{2}} &
\qquad\hat a = a/L \qquad
\hat{m}= L\, {\rm m}\, e^{-3\phi_0/\sqrt{2}} \,.
\end{aligned}
\label{eq:LifshitzRescalings}
\ee
Equations (\ref{6deom1}) to (\ref{6deom5}) then reduce to a simple set of 
algebraic equations with the following general solutions
\be
\begin{aligned}
&\hat b^2 = z-1 \qquad \hat g^2= 2 z(4+z) &
& \frac{\hat{m}^2}{2}=\frac{6+z \mp 2\sqrt{2(z+4)}}{z} 
\\
&\hat q^2=\frac{ (2+z) (z-3)\pm 2 \sqrt{2(z+4)}}{2 z} & 
& \frac{1}{\hat a^2} = 6+3z \mp 2\sqrt{2(z+4)} \, .
\end{aligned}
\label{eq:lifsolns}
\ee
These define two families of Lifshitz spacetimes, one for each branch of the
square root in (\ref{eq:lifsolns}). 
The requirement that $\hat b$ is real restricts $z$ to $z\geq1$ and 
for the lower sign choice, for $\hat q$ to be real one finds that $z$ 
must be greater than approximately $4.29$.

In addition to these Lifshitz solutions, the system also allows for 
an independent  
one parameter family of adS solutions when  $z=1$ and $\hat b=0$.  
These solutions can be parametrized by either $\hat g$ or $\hat m$,
the latter case giving
\be
{\hat g}=\frac{{\hat m}^2+6}{2\hat m}\,,\;\;\;
{\hat q}^2=-\frac{5 {\hat m}^4 - 36{\hat m}^2 +36}{16\hat m^2}\,, \;\;\;
\frac{1}{\hat a^2}=\frac{5{\hat m}^4- 12{\hat m}^2+ 36}{8\hat m^2}\,.
\label{eq:adssols}
\ee
In this case the requirement that $\hat q$ is real implies that 
$\hat m\in[\sqrt{\frac{6}{5}},\sqrt{6}]$. When  $\hat m\,=\,\sqrt{10}-2$ 
the adS and (upper sign) Lifshitz solutions touch at $z=1$. 
Figure \ref{fig:liadssp} represents the values of $z$ and $\hat m$ that these 
Lifshitz and adS solutions can take.
\FIGURE{
\epsfig{file=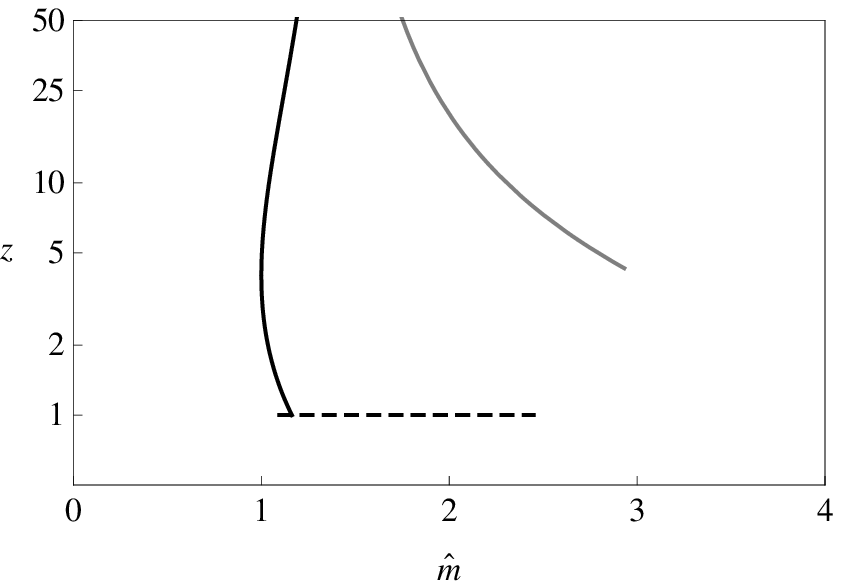,width=0.6\textwidth}
\caption{Plot showing the values of $z$ and $\hat m$ that the Lifshitz and adS 
solutions can take.  The horizontal dashed line indicates the adS solutions, 
with $z=1$. The black line corresponds to the Lifshitz solutions with the 
upper sign choice in (\ref{eq:lifsolns}) and the grey to the lower 
sign choice. Notice that adS and Lifshitz solutions meet 
at $\hat m\,=\,\sqrt{10}-2$. }
\label{fig:liadssp}
}

In what follows, we continue to analyze adS and 
Lifshitz solutions, generalizing the previous
discussion to the case of asymptotically adS and Lifshitz space-times. 
Insights acquired when analyzing asymptotically adS configurations 
will be of great help when considering  asymptotically Lifshitz space-times. 

\subsection{General planar spacetimes}

Our main aim  is to characterize  asymptotically adS and  
Lifshitz black hole solutions.
We look for solutions which respect the planar symmetry 
and static nature of the metric (\ref{eq:6dMetric}), meaning the alterations 
need only have radial dependence, 
$\phi=\phi(r)$ and
\be
ds^2 = L^2 \left [ e^{2f(r)} dt^2 - e^{2c(r)} d{\bf x}^2 - e^{2d(r)} dr^2
\right ]-e^{2h(r)} dH_2^2\,.
\label{genplang}
\ee
We choose our field strength Ans\"atze to be
\be
F^{(3)}_{tr}=L^2Q(r)\,, \;\;\; F^{(3)}_{y_1 y_2}=\frac{q}{y_2^2}\,,\;\;\;
B_{x_1x_2}=\frac{L^2}{2}e^{-\sqrt{2}\phi_0}\,  P(r)\,,
\ee
which gives the gauge equations:
\bea
\left ( e^{2c +2h -f -d -\sqrt{2}\phi } Q  \right )' &=& 
q e^{-\sqrt{2}\phi_0}P' \label{Feqn} \\
\left ( e^{f -d +2h -2c +2\sqrt{2}\phi }e^{-\sqrt{2}\phi_0}P'
\right )' &=& m^2 L^2 e^{-\sqrt{2}\phi_0}P e^{f+d-2c+2h-\sqrt{2}\phi} 
+ 4L^2 q Q \,. \label{Heqn}
\eea
Integrating (\ref{Feqn}) and noting that $Q\to0$ as $P\to0$ 
from (\ref{Heqn}), we obtain:
\be
Q(r)= e^{\sqrt{2}\phi(r)+f(r)+d(r)-2c(r)-2h(r)}qe^{-\sqrt{2}\phi_0}P(r)\,,
\ee
hence there is a single equation of motion for the gauge fields:
\be
\left( e^{f-d+2h-2c+2\sqrt{2}\phi} P'\right)' =
L^2 P \, e^{f+d-2c+2h-\sqrt{2}\phi} \left( m^2 
+ 4q^2 e^{2\sqrt{2}\phi-4h} \right)\,. \label{eq1:gauge}
\ee
The remaining equations are:
\bea
\frac{\sqrt{2}}{\sqrt{g}} \left( \frac{\sqrt{g}\phi'}{L^2 e^{2d}} \right )' &=&
P^2 e^{-2\sqrt{2}\phi_0 -\sqrt{2}\phi -4c} 
\left (q^2e^{2\sqrt{2}\phi-4h} -\frac{\rm m^2}{4} \right )
+ \frac{P^{\prime2}}{2L^2} e^{2\sqrt{2}(\phi-\phi_0)-4c-2d} 
\nonumber \\
&&-q^2 e^{-\sqrt{2}\phi - 4h} 
- \frac{1}{4} \left ( g^2 e^{\sqrt{2}\phi} - 4{\rm m} g e^{-\sqrt{2}\phi}
+3{\rm m}^2 e^{-3\sqrt{2}\phi} \right ) \label{eq1:phi}\\
\frac{2}{\sqrt{g}} \left ( \frac{\sqrt{g} f'}{L^2 e^{2d}} \right )' &=&
P^2 e^{-2\sqrt{2}\phi_0 -\sqrt{2}\phi -4c} 
\left (3q^2e^{2\sqrt{2}\phi-4h} +\frac{\rm m^2}{4} \right )
+ \frac{P^{\prime2}}{2L^2} e^{2\sqrt{2}(\phi-\phi_0)-4c-2d} 
\nonumber \\
&&+ q^2 e^{-\sqrt{2}\phi - 4h} 
+\frac{1}{4} \left ( g^2 e^{\sqrt{2}\phi} + 4{\rm m}g e^{-\sqrt{2}\phi}
-{\rm m}^2 e^{-3\sqrt{2}\phi} \right )\\
\frac{2}{\sqrt{g}} \left ( \frac{\sqrt{g} c'}{L^2 e^{2d}} \right )' &=&
P^2 e^{-2\sqrt{2}\phi_0 -\sqrt{2}\phi -4c} 
\left (-q^2e^{2\sqrt{2}\phi-4h} -\frac{3 {\rm m}^2}{4} \right )
- \frac{P^{\prime2}}{2L^2} e^{2\sqrt{2}(\phi-\phi_0)-4c-2d} 
\nonumber \\
&&+ q^2 e^{-\sqrt{2}\phi - 4h} 
+\frac{1}{4} \left ( g^2 e^{\sqrt{2}\phi} + 4{\rm m}g e^{-\sqrt{2}\phi}
-{\rm m}^2 e^{-3\sqrt{2}\phi} \right )\\
\frac{2}{\sqrt{g}} \left ( \frac{\sqrt{g} h'}{L^2 e^{2d}} \right )' &=&
P^2 e^{-2\sqrt{2}\phi_0 -\sqrt{2}\phi -4c} 
\left (-q^2e^{2\sqrt{2}\phi-4h} +\frac{\rm m^2}{4} \right )
+ \frac{P^{\prime2}}{2L^2} e^{2\sqrt{2}(\phi-\phi_0)-4c-2d} 
\nonumber \\
&&-3q^2 e^{-\sqrt{2}\phi - 4h} 
+ \frac{1}{4} \left ( g^2 e^{\sqrt{2}\phi} + 4{\rm m}g e^{-\sqrt{2}\phi}
-{\rm m}^2 e^{-3\sqrt{2}\phi} \right ) - 2e^{-2h} 
\,,\;\;\;\;\;\;\;\;\;\;\label{eq1:P}
\eea
together with the first integral of the Einstein equations:
\bea
&& 2f'c' + 2f'h' + 4c'h' + c^{\prime2}
+h^{\prime2} - \phi^{\prime2}
- \frac{P^{\prime2}}{4} e^{2\sqrt{2}(\phi-\phi_0)-4c} = \label{BI}
\\
&&{L^2e^{2d}} \left[ -e^{-2h}
- q^2 e^{-\sqrt{2}\phi - 4h}-P^2 e^{-2\sqrt{2}\phi_0 -\sqrt{2}\phi -4c} 
\left (q^2e^{2\sqrt{2}\phi-4h} +\frac{\rm m^2}{4} \right )
+ V(\phi) \right] \,.\nonumber 
\eea

In analysing the solutions of these equations, it is particularly useful 
to consider the equations of motion from a dynamical systems perspective. 
The exact Lifshitz or adS geometries correspond to fixed points of the 
dynamical system, and a perturbative analysis around the fixed points indicates
the flows of general interior solutions to the asymptotic Lifshitz or
adS geometry. This general methodology was used in \cite{BGR} to explore
flows between exact Lifshitz and adS solutions.

Although the system 
(\ref{eq1:gauge}-\ref{eq1:P}) appears to be nine
dimensional, there is a redundant gauge degree of freedom corresponding
to a rescaling of the coordinates, and also the Bianchi identity 
(\ref{BI}),
which reduces the order of the system to seven. 
Perturbing around a fixed point solution therefore will give a
seven dimensional solution space 
(some of which will
correspond to unphysical singular solutions), spanned by the eigenvectors of 
the perturbation operator around the critical point, with a radial
fall-off given by the corresponding eigenvalue $\Delta$.
In  appendix \ref{appds}  we provide the details of a dynamical 
systems analysis of these equations of motion, and the derivation 
of the eigenvalues. The eigenvalues will provide crucial information 
about the physical charges that characterise a given solution.  

\section{Analytic results}

In order to develop a general understanding that will be later 
used to determine numerical black holes configurations, 
it is useful to  analytically explore the allowed asymptotic behaviour of 
adS and Lifshitz solutions for our theory.  
In this section, by implementing the dynamical system analysis 
developed in appendix \ref{appds},  
we will first identify perturbative solutions which describe flows 
towards an adS boundary at large $r$.  Among these is the exact 
adS-Schwarzschild solution, and we analyse linearized solutions 
about this background, throughout the space-time from horizon 
to boundary.  We then study flows towards a Lifshitz boundary 
at large $r$, for arbitrary dynamical exponent $z$.  In this way, 
we are also able to observe how asymptotically Lifshitz geometries 
for $z>1$ reduce to asymptotically adS geometries at $z=1$. 

For convenience, we choose the radial coordinate to correspond to the
area gauge as in (\ref{eq:LifshitzMetric}), i.e. $c(r)=\log r$, 
and rewrite our metric and 
scalar functions in terms of deviations from the known Lifshitz background:
\be
\begin{aligned}
&\phi(r) = \phi_0 + \varphi(r) /\sqrt{2}\,,\;\;\; h(r)  
= \log a+ \half \ln H(r)\,,  \;\;\;  P(r) = r^2 p(r) \\
\,\;\;\;
& f(r)  = z \ln r + \half \ln F(r)  \,, \;\;\;
d(r)  = -\ln r - \half \ln D(r) \,.
\end{aligned}
\ee
This gives us the field equations
\bea
\frac{(r^{z+3}\sqrt{FD}\,H\varphi' )'}{r^{z+1}H\sqrt{F/D}}
&=& - \frac{{\hat q}^2 e^{-\varphi}}{H^2}
+ p^2  \left (\frac{{\hat q}^2 e^{\varphi}}{H^2}
-\frac{{\hat m}^2 e^{-\varphi}}{4} \right )
+ 2 D e^{2\varphi} \left ( p + \frac{r p'}{2} \right )^2
-\frac{\partial{\hat V}}{\partial\varphi} \label{eq:phi}\\
\frac{(r^{z+3}\sqrt{FD}\, H')'}{r^{z+1}H\sqrt{F/D}} &=&
-3\frac{{\hat q}^2 e^{-\varphi}}{H^2}
- p^2  \left (\frac{{\hat q}^2 e^{\varphi}}{H^2}
-\frac{{\hat m}^2 e^{-\varphi}}{4} \right )
+ 2 D e^{2\varphi} \left ( p + \frac{r p'}{2} \right )^2
\hskip 25mm \nonumber  \\    
&&\qquad+ {\hat V}(\varphi) - \frac{2}{{\hat a}^2 H}  \label{eq:H}\\
\frac{(r^{z+3}\sqrt{D/F}\,HF')'}{r^{z+1}H\sqrt{F/D}} &=&
(1-z) \frac{{\hat q}^2 e^{-\varphi}}{H^2}
+ p^2  \left (\frac{(3+z){\hat q}^2 e^{\varphi}}{H^2}
+\frac{(1+3z){\hat m}^2 e^{-\varphi}}{4} \right )
\nonumber \\
&& \qquad + 2 (1+z) D e^{2\varphi} \left ( p + \frac{r p'}{2} \right )^2
+ (1-z) {\hat V}(\varphi) \label{eq:F}\\
\frac{(2r^{z+2}H\sqrt{FD}\,)'}{r^{z+1}H\sqrt{F/D}} &=&
\frac{{\hat q}^2 e^{-\varphi}}{H^2}
- p^2  \left (\frac{{\hat q}^2 e^{\varphi}}{H^2}
+\frac{3{\hat m}^2 e^{-\varphi}}{4} \right )
- 2 D e^{2\varphi} \left ( p + \frac{r p'}{2} \right )^2
+ {\hat V}(\varphi)\,, \label{eq:D}
\eea
together with the gauge equation
\be
\left (r^z\sqrt{FD}\,e^{2\varphi} H  \left ( 2 p + r p' \right )\right)'
= r^{z-1} p He^{-\varphi} \sqrt{F/D} \left ( {\hat m}^2 
+ \frac{4{\hat q}^2 e^{2\varphi}}{H^2} \right )\,,  \label{eq:P}
\ee
and the first integral
\bea
D \left [ 2z+1+\frac{rF'}{F} + (z+2) \frac{rH'}{H} +\frac{r^2 H'F'}{2HF}
+\frac{r^2H^{\prime2}}{4H^2} - \frac{r^2\varphi^{\prime2}}{2}
-\left ( p + \frac{r p'}{2}\right )^2 e^{2\varphi} \right ] \nonumber\\
= {\hat V}(\varphi) - \frac{1}{{\hat a}^2H} - \frac{{\hat q}^2e^{-\varphi}}{H^2}
-p^2 \left ( \frac{{\hat q}^2e^\varphi}{H^2}+\frac{{\hat m}^2e^{-\varphi}}{4}
\right )\,, \hskip 1cm
\eea
where 
\be
{\hat V}(\varphi) = \frac14 \left ( {\hat g}^2 e^\varphi
+4{\hat m}{\hat g} e^{-\varphi} - {\hat m}^2 
e^{-3\varphi} \right )\,.
\ee

\subsection{Anti-de Sitter solutions}

We  start with a detailed  analysis of  
the adS branch of solutions to our system. These
configurations are easier to study than Lifshitz space-times, yet
they give insights into what charges are inherent in the system,
and what aspects of black hole solutions are uniquely Lifshitz.
In the adS case, of course, we already know an analytic black hole, 
the adS-Schwarzschild solution:
\be
H = 1\,,\;\;\; \varphi = p = 0\,,\;\;\; 
F = D = 1 - \left ( \frac{r_+}{r} \right )^3 \,.
\label{eq:adssch}
\ee
From the point of view of the dynamical system discussed in 
appendix \ref{appds}, this corresponds to the nonlinear evolution 
into the interior of the asymptotic eigenvalue\footnote{Asymptotic 
eigenvalues $\Delta_i$ control
the fall-off $r^{-\Delta_i}$ of a given function at large $r$.} $\Delta = -3$  
(the $1/r^3$ fall-off), which has an eigenvector with components only 
in the directions corresponding to deformations of the 4D geometry,
as in eq.~(\ref{eq:adssch}).

It is useful however to continue with a more general analysis of 
perturbative solutions which asymptote adS, as this will  
enable a more direct comparison with the Lifshitz case.
A general analysis of the perturbations around the adS fixed 
point (see appendix \ref{appds}) yields the eigenvalues 
plotted in figure \ref{fig:Adseval}.
As explained in detail in the figure caption,
these eigenvalues give the exponents of $r$ in the asymptotic 
solutions for the various fields, as function of $\hat m$. Each field has a 
fall-off of a pair of exponents which are symmetric about $-3/2$,
and whose coefficients can be interpreted as a source and operator in the 
boundary field theory. The pure `black hole' mode, $\Delta = -3$ is in this
sense dual to the zero mode which takes us along the parameter
space of adS solutions.
\FIGURE{
\includegraphics[width=10cm]{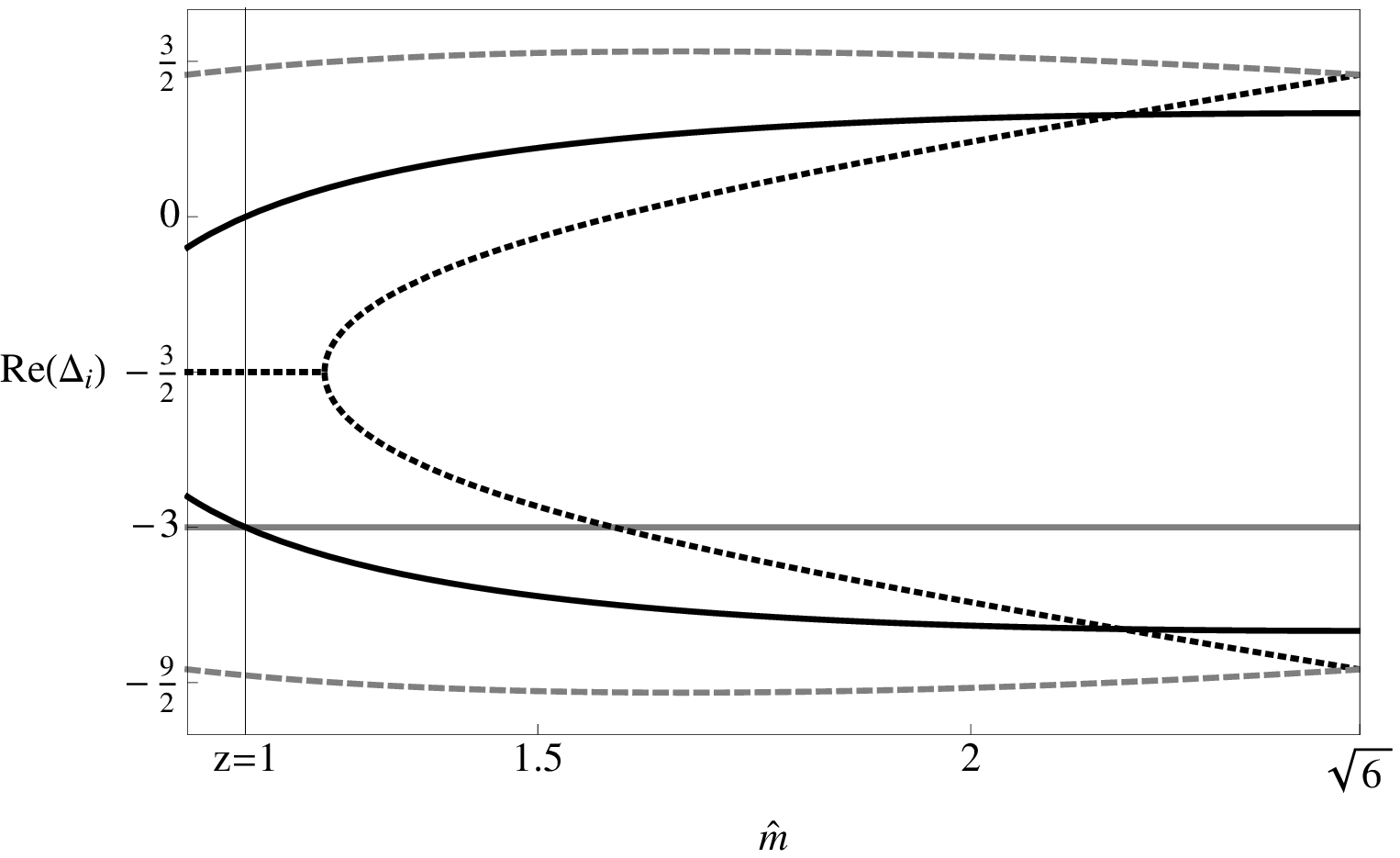}
\label{fig:Adseval}
\caption{Plot showing the real part of the field exponents 
as they asymptotically approach an adS space-time, as a function 
of the parameter $\hat m$.  Each pair of exponents sums to $-3$. The solid 
black lines correspond to switching on a B-charge in the interior 
spacetime, and the solid horizontal grey line is the black hole solution. 
A combination of the dotted and dashed black and grey lines corresponds
to switching on the dilaton and breather modes. (The joining of the dotted 
black lines for small $\hat m$ indicates that the exponents turn complex.) 
Notice that at ${\hat m} = \sqrt{10} -2$, which is the value of $\hat m$ at
which the adS fixed point solution is equivalent to a $z=1$ upper 
branch Lifshitz solution, there is a degeneracy in the eigenvalues: 
the mass and charge deformations have the same fall-off.
}}

A combination of the dotted and dashed black and grey lines 
in figure \ref{fig:Adseval}  corresponds
to switching on the dilaton and breather modes, keeping $P$ equal to zero.
The solid black lines in the figure are associated with the  
turning on of the 2-form charge $P$ only. 
At ${\hat m} = \sqrt{10}-2$, the value of $\hat m$ for which the pure adS
and Lifshitz solutions coincide (since $z=1$), and the charge 
and mass deformations of 
adS become degenerate. Since the $P$-equation decouples at leading
order, it is not difficult to extract the charged perturbations of adS: 
\be
H = F =D = 1\,,\;\;\; \varphi = 0\,,\;\;\; 
p= p_0 - \frac{p_3}{r^3} \,.
\label{eq:adscharg}
\ee
The $p_0$ deformation corresponds to the zero mode (the upper solid black line 
at ${\hat m} = \sqrt{10}-2$ in figure \ref{fig:Adseval}) which moves
the solution onto the Lifshitz branch  whereas the $p_3$ 
deformation identifies the pure charge eigenvector.

As we learn in what follows, perturbations of asymptotically 
Lifshitz configurations have different eigenvectors and eigenvalues, 
conveniently parameterized by the quantity $z$. On the other hand, 
when ${\hat m} = \sqrt{10}-2$ 
(corresponding to $z=1$) the Lifshitz deformations and adS deformations
should coincide, as there is a degeneracy of the eigensystem 
at this point: the corresponding Lifshitz deformations are a 
combination of the pure mass and pure charge asymptotically adS solutions.
In Section \ref{sec:lifshitzexact} we analyze the corresponding 
Lifshitz asymptotic solutions and how they join the adS ones at  $z=1$. 

Turning now to black hole solutions, we next explore linearized
solutions for scalar and gauge charges around the known
adS-Schwarzschild black hole background of eq.~(\ref{eq:adssch}).
We stress that the eigenvalue analysis discussed above 
refers to perturbations of the full system around the adS background, 
and as such  describe how the geometry and fields asymptote the
adS boundary at large $r$. In contrast, in the next two subsections
we seek linearized solutions of
either the $B-$field, or the dilaton and breather mode, 
around the black hole background. These, 
at leading order,  do {\it not} include perturbations of the black hole
geometry, but they are solutions of the scalar or gauge fields for
the full range of the space-time from the horizon to the boundary.
Our linearized solutions, therefore, should asymptote one or more
of the eigenvalue solutions 
near infinity, at least for the fields we are perturbing. 
On the other hand, they are more informative since they provide 
some further understanding on how   
field configurations  behave  in the black hole background. 
Our aim here is to identify the eigensolutions of the full
system, so as to have a physical interpretation of the various eigenvectors
which we can then use to understand the Lifshitz system.

\subsubsection{$B-$charge}

We start by allowing $p$ to vary, but keeping  $\varphi$ and $H$ constant
over the black hole background of eq.~(\ref{eq:adssch}).  
Since $P$ is related to the electric field and the flux of 
the $B_{AB}$ field of the system we shall refer to these solutions 
as {\it charged} black holes.  Note that, to leading order, 
the $p-$equation (\ref{eq:P}) decouples from the other equations 
and around a black hole background is
\be
\left [ \left ( 1 - \frac{r_+^3}{r^3} \right ) (r^2 p)' \right ]'
= \left ( {\hat m}^2 + 4 {\hat q}^2 \right ) p
= \left ( \frac{36{\hat m}^2-36-{\hat m}^4}{4{\hat m}^2}  \right ) p\,.
\ee
Writing $x = (r_+/r)^3$ and definining
\be
\nu_\pm = (\pm \sqrt{(36-{\hat m}^2)({\hat m}^2-1)} - {\hat m})/6{\hat m}\,,
\ee
we obtain  a closed  solution for $P=r^2 p$ in terms of hypergeometric functions:
\bea
P(r) &=& \Gamma[2\nu_-+4/3] \,\Gamma[\nu_++4/3]\, 
\Gamma[\nu_+]\, x^{\nu_+}
\,_2F_1 [ \nu_+,\nu_+ +4/3,2\nu_+ +4/3;x(r)]  \nonumber\\
&&-
\Gamma[2\nu_+ +4/3] \,\Gamma[\nu_- +4/3]\, \Gamma[\nu_-] \,x^{\nu_-}
\,_2F_1 [ \nu_-,\nu_- +4/3,2\nu_- +4/3;x(r)]\,, \nonumber\\ \;\;\;\;\;
\label{phypergeo}
\eea
where the constants are chosen to give a nonsingular combination at
$x=1$, the position of the horizon. A quick glance at the eigenvalue plot,
figure \ref{fig:Adseval}, shows that, unless ${\hat m} < \sqrt{10}-2$, the
$\nu_+$ hypergeometric function will blow up at infinity. Thus for 
nonsingular linearised solutions we take ${\hat m} < \sqrt{10}-2$
(this does not mean that charged black holes do not exist 
for ${\hat m} > \sqrt{10}-2$, simply that they have strong
gravitational backreaction).
Figure \ref{fig:Ancharge} (on the left) shows a representative 
sample charged $B-$field around an adS black hole with ${\hat g}^2 = 52/5$.  
We will further numerically  explore these black holes in section \ref{NUMBH}.
\FIGURE{
\includegraphics[width=7.25cm]{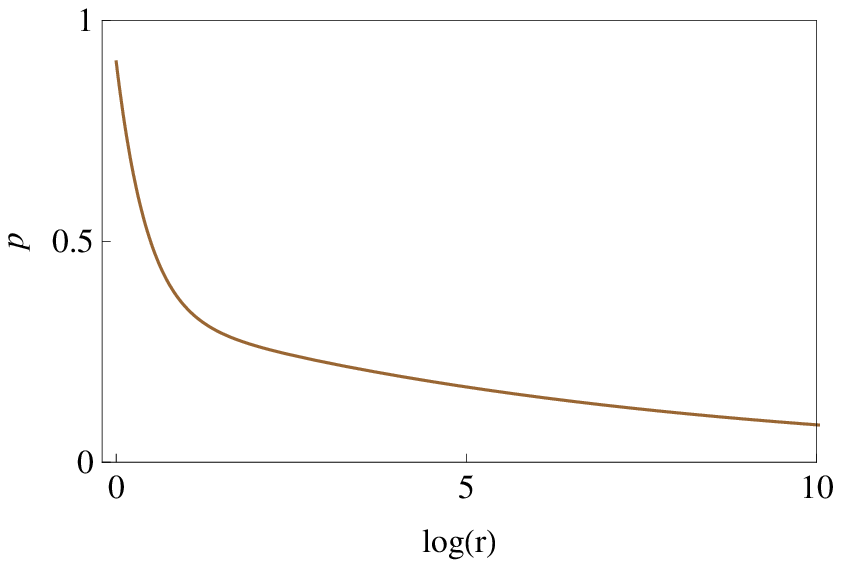}
\includegraphics[width=7.25cm]{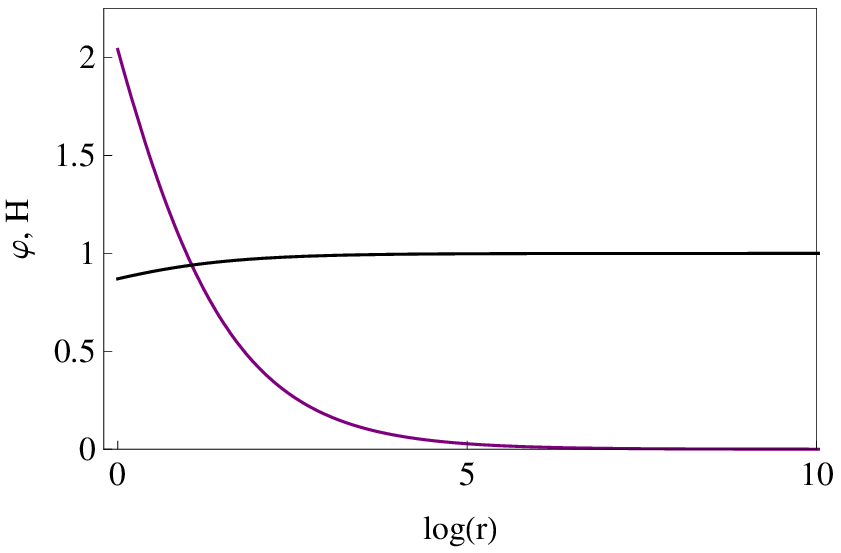}
\caption{The left plot shows a perturbation of $p$ for $\hat m=\sqrt{10}-2$ 
whose asymptotic behaviour is consistent with (\ref{eq:adssch}).  
The right plot shows perturbations of $\varphi$ (in purple) and $H$ 
(in black) for $\hat m=3/2$, which are consistent with (\ref{eq:adssch}).}
\label{fig:Ancharge}
}

\subsubsection{Scalar charge}

From the perspective of the 4D geometry, the breather
mode of the hyperbolic geometry appears as a 4D scalar,
and indeed the scalar equations, (\ref{eq:phi}, \ref{eq:H}) 
are independent of perturbations of the geometry at linear
order. Extracting these scalar equations gives a second order system:
\be
{\cal L} \left(\begin{matrix} \delta \varphi \\ \delta H \end{matrix}\right )
= \left [ \begin{matrix} (3{\hat m}^2 - {\hat g}^2)/2 & 2{\hat q}^2\\
2 {\hat q}^2 & 2({\hat q}^2 + 3) \end{matrix} \right ]
\left(\begin{matrix} \delta \varphi \\ \delta H  \end{matrix}\right ) \,,
\label{scalred}
\ee
where ${\cal L}$ is the linear operator
\be
{\cal L}\, X\, = \,\frac{1}{r^2} \frac{d\ }{dr} \left [  r^4
\left ( 1 - \frac{r_+^3}{r^3} \right ) \,\frac{dX}{dr} \right ]\,.
\ee
Diagonalising the matrix on the RHS of (\ref{scalred}) yields
two eigenvalues and eigenvectors which correspond to the two 
pairs of exponents indicated by the grey dashed and black 
dotted lines in figure \ref{fig:Adseval}. The fully coupled system 
also has perturbations of the geometry, although the asymptotic
exponents indicated in figure \ref{fig:Adseval} represent the
fall-off at large $r$. The eigenvalues of (\ref{scalred}), however,
are linearized solutions around the given black hole background for
all $r$. Clearly, examining the large $r$ behaviour from figure
\ref{fig:Adseval} shows that the dotted grey branch cannot yield a solution 
which is regular at both horizon and infinity: hence these branches have
significant backreaction on the geometry. However, for 
${\hat m}<(6-\sqrt{6})/\sqrt{5}$ the black dotted branch gives 
the regular solution:
\be
\left(\begin{matrix} \delta \varphi \\ \delta H \end{matrix}\right )
= \left ( \begin{matrix} -8{\hat m}^2
\pm\sqrt{36-60 {\hat m}^2+89 {\hat m}^4}\\
5{\hat m}^2 - 6 \end{matrix}\right ) 
X \left [ \left ( \frac{r_+}{r} \right )^3\right] \,,
\ee
where
\be
X[x] = 
\Gamma[2\mu_-] \Gamma[\mu_+]^2 x^{\mu_+} \,_2F_1 [ \mu_+,\mu_+,2\mu_+;x] -
\Gamma[2\mu_+] \Gamma[\mu_-]^2 x^{\mu_-} \,_2F_1 [ \mu_-,\mu_-,2\mu_-;x]
\label{schypergeo}
\ee
and 
\be
\mu_\pm = \half \left [ 1 \pm \sqrt{1 + 4\lambda/9} \right ]
\ee
are given in terms of the eigenvalue
\be
\lambda=\frac{1}{8 \hat m^2}\left(3 \hat m^4+36 \hat m^2-36 + \left(\hat 
m^2-6\right) \sqrt{36-60 \hat m^2+89 \hat m^4}\right) \,.\label{eq:lambda}
\ee

As is noted in \cite{BGR} the straight part of the dotted black 
curve, where $\sqrt{6/5}<\hat m<1.254$, indicates an imaginary exponent.
This occurs when $\lambda<-9/4$ in (\ref{eq:lambda}) and
is analogous to a mass violating the Breitenlohner-Freedman bound,
\cite{BF}. This fact is indicative of a possible instability,
most likely a flow from one adS branch to another.  
Figure \ref{fig:Ancharge} (on the right) shows the field profiles 
for the representative value ${\hat m}=3/2$. 

\subsection{Lifshitz solutions}\label{sec:lifshitzexact}

We now apply to Lifshitz configurations the same techniques we developed to  
characterize adS solutions. As we will see, the intuition we developed 
for adS will help in characterizing the richer Lifshitz 
configurations. Unlike the adS case, there is no straightforward 
analytic Lifshitz black hole solution for our system\footnote{See 
however appendix \ref{appel} for examples of exact analytic solutions 
in similar dilatonic theories.}. That this will be the case can be seen by 
analyzing  the Lifshitz fixed point, where all of the eigenvectors 
corresponding to perturbations around the asymptotic Lifshitz geometry 
generically have components in every field. In this instance, 
there is no pure geometric deformation to the Lifshitz space, and 
any deformation necessarily includes a scalar and gauge profile.
\FIGURE{
\includegraphics[width=7.25cm]{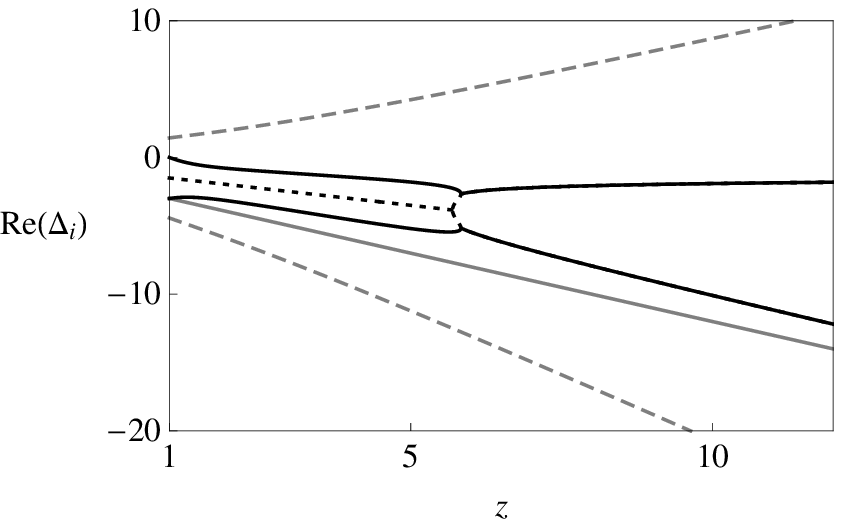}
\includegraphics[width=7.45cm]{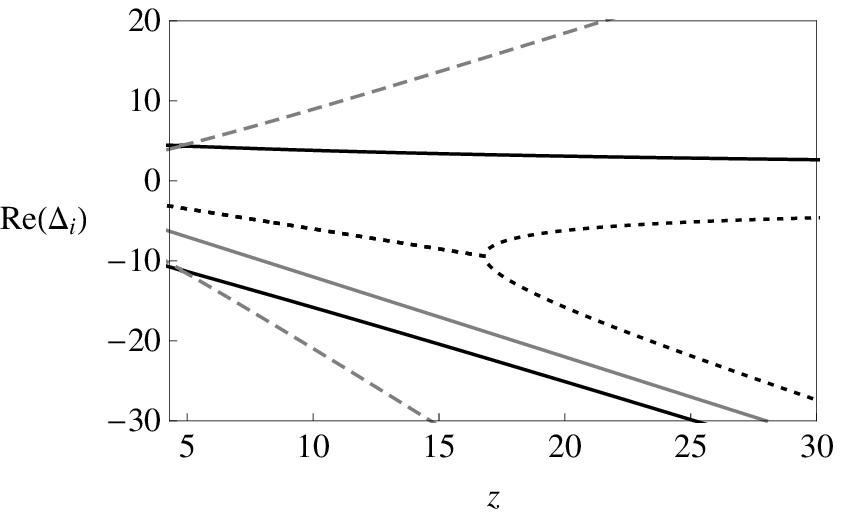}
\caption{Plots of the real parts of the eigenvalues of perturbations around
the Lifshitz solution as a function of $z$.  The left and right plots 
correspond to the upper and lower sign choices in (\ref{eq:lifsolns}) 
respectively. For the upper sign choice, the eigenvalues are all real 
only in the region $5.69 < z < 5.83$, whereas for the lower sign choice 
one finds real eigenvalues for $z > 16.82$.}
\label{fig:liffp}
}

The analysis of the Lifshitz point is given in appendix \ref{appds}, and the 
system of eigenvalues plotted in figure \ref{fig:liffp}. Comparing this plot to
the adS case, we see that the eigenvalues are symmetric around $-(z+2)/2$,
and in particular, an eigenvalue $\Delta = -(z+2)$ exists, which is continuous
with the pure black hole adS eigenvalue, $\Delta=-3$. Indeed, plotting the
Lifshitz and adS eigenvalues side by side shows how the perturbations around the
critical points merge as $z\to1$ or ${\hat m} \to \sqrt{10}-2$. Using the
intuition obtained from analysing the adS solutions, one might
expect that the mass and charge perturbations continue from
the adS side into the Lifshitz side. However, things are not 
so straightforward, and are actually more interesting.
While the subspace spanned by the two eigenvectors clearly is the same on each
solution branch, the eigenvector basis need not be: the adS eigenvectors
are either pure charge or pure geometry, whereas the Lifshitz perturbations
include all fields. A careful tracking of the perturbations as $z\to1$ 
indicates that the Lifshitz perturbations can be thought as 
corresponding  to a $\pi/4$ rotation of the adS perturbations. 
Thus, the ``$-(z+2)$" eigenvector is actually a charged black hole, 
most likely an extremal black hole given the combination of adS eigenvectors.

It is worth exploring in more detail the analytic expansions of
the functions, especially for the special cases identified above.
First of all, by analysing the linear perturbations for the
eigenvalue $\Delta=-(z+2)$, we find the solution (for general $z$):
\bea
\delta \varphi &=& \frac{2\, M\, \sqrt{z-1}}{r^{z+2}}
\Big[ \sqrt{2(4+z)} 
\left(-48 - 8 z + 14 z^2 + 4 z^3\right) \nonumber\\
&&\hskip2cm+\, \left(136 + 40 z - 40 z^2 - 21 z^3 - 2 z^4\right)
\Big] \nonumber\\
F&=&1-\frac{2 M \sqrt{z-1}}{(2+z)\,r^{z+2}}\,  \Big[\sqrt{2(4+z)}
\left(96 - 352 z - 228 z^2 - 22 z^3 + 4 z^4\right) 
\nonumber\\
&&\hskip1.5cm +\,\left(-272 + 952 z + 760 z^2 
+ 146 z^3 - 11 z^4 - 11 z^5 - 2 z^6\right)\Big] \nonumber\\
D&=&1 + \frac{2\, M\, \sqrt{z-1}}{r^{z+2}}\Big[ \sqrt{2(4+z)} 
\left(-48 + 40 z + 38 z^2 + 4 z^3\right) \nonumber\\
&&\hskip2.5cm+\, \left(136 - 96 z - 124 z^2 - 19 z^3 + 7 z^4 + 2 z^5
\right)\Big] \nonumber\\
H&=&1 + \frac{2M\, \sqrt{z-1}}{r^{z+2}}\Big[ \sqrt{2(4+z)} 
\left(-48 + 8 z + 2 z^2\right) + \left(136 - 8 z - 12 z^2 + 5 z^3 + 2 z^4 \right)\Big] \nonumber\\
p&=&\sqrt{z-1}+ \frac{2M}{r^{z+2}}\Big[ \sqrt{2(4+z)} \, 
\left(-48 - 8 z + 22 z^2 + 6 z^3\right) \nonumber\\
&&  \hskip1.5cm+\left(136 + 40 z - 64 z^2 - 23 z^3 + 7 z^4 + 2 z^5\right)\Big]\label{pertzp2}
\eea
where $M$ is some integration parameter.  This solution is valid at 
first order in perturbations around the pure Lifshitz solution with 
arbitrary dynamical exponent $z$, and makes manifest that all fields
are normally switched on for Lifshitz configurations. 
We additionally  checked that the system of equations can be solved also
at next to leading order, providing corrections to the 
above profiles that scale as $M^2/ r^{2(z + 2)}$: the 
expression for the $z-$dependent coefficients is however too 
long to be presented here.  Notice that  the limit $z\to 1$ is
well behaved, and leads apparently to the perturbative pure charge adS 
solution we discussed in equation (\ref{eq:adscharg}). 

We can also characterize the solution corresponding to the black line 
in the left panel of figure  \ref{fig:liffp}, that joins with the 
grey line when $z\to 1$. In order to do this, we expand the eigenvalues  
tending to the $\Delta=-3$ near $z\to 1$ to obtain:
\bea
\Delta_1&=&- 3- (z-1)\\
\Delta_2&=&- 3+\frac{1}{189} (260 \sqrt{10}-701) (z-1)
+ {\cal O}\left[(z-1)^2 \right]
\eea
The first is just the eigenvalue $\Delta = -z-2$ rewritten as an
expansion around $z=1$, with a corresponding solution, (\ref{pertzp2}),
that can also be expanded near $z=1$. The second eigenvalue, $\Delta_2$, 
is instead that of the black curve in figure \ref{fig:liffp}, and we
can similarly determine its corresponding solution. To leading
order in $z-1$ we find:
\bea
\delta \varphi_1 &=& \frac{\mu\, \sqrt{z-1}}{126\,r^{\Delta_1}}
(31 - 40 \sqrt{10})  \hskip1cm,\hskip1cm \delta \varphi_2 \,
=\, \frac{\mu\, \sqrt{z-1}}{126\,r^{\Delta_2}}
(31 - 40 \sqrt{10}) \nonumber\\
F_1&=&1-\frac{\mu \sqrt{z-1}}{63\,r^{\Delta_1}}\,  ( 65 \sqrt{10} -149)
\hskip0.2cm,\hskip1cm F_2 \,=\,1- \frac{\mu\, \sqrt{z-1}}{3\,r^{\Delta_2}}
\nonumber\\
D_1&=&1 + \frac{\mu \, \sqrt{z-1}}{63 \,r^{\Delta_1}}  (11 + 25 \sqrt{10}) 
\hskip0.4cm,\hskip1cm D_2 \,=\,1+ \frac{\mu\, \sqrt{z-1}}{63\,r^{\Delta_2}} 
(139 - 40 \sqrt{10})
\nonumber\\
H_1&=&1 + \frac{\mu\, \sqrt{z-1}}{126 \,r^{\Delta_1}} 
(101 - 20 \sqrt{10})
\hskip0.2cm,\hskip1cm H_2 \,=\,1 + \frac{\mu\, \sqrt{z-1}}{126 \,r^{\Delta_2}} 
(101 - 20 \sqrt{10}) \nonumber\\
p_1&=&\sqrt{z-1}+ \frac{\mu}{r^{\Delta_1}} \hskip2.5cm,\hskip1cm 
p_2\,=\,\sqrt{z-1}+ \frac{\mu}{r^{\Delta_2}} \label{pertzp3}
\eea
for some integration constant $\mu$. Notice that both these eigenvector solutions
approach the pure charge adS solution for in the limit $z\to 1$. 
On the other hand, for small values of $(z-1)$ they differ 
(by an identical amount) {\it only} in the metric components $F$ and $D$ (up 
to corrections suppressed by powers of $\sqrt{z-1}$). 
The difference in the eigenvectors is consequently  due  to contributions 
of pure geometry  to the Lifshitz configuration, that we know corresponds
to the adS-Schwarzschild eigenvector in the pure adS case. In this sense, 
we can regard the two  eigenvectors (\ref{pertzp3}), in the limit of 
small $(z-1)$, as if each forming a $\pi/4$ degree angle with the 
eigenvectors of asymptotically adS configurations. 

\section{Numerical Black Hole Solutions}
\label{NUMBH}

Having developed an analytical understanding of the asymptotic 
properties of the solutions of the system,
we now present some numerical solutions to the 
fully coupled system of field equations.  
In this section, we begin by computing and analysing adS black hole 
solutions, followed by their Lifshitz generalizations,   
and then uplift the solutions to type IIA supergravity.

To obtain black hole solutions for our system we must ensure 
that our boundary conditions are consistent with the nature of the near horizon 
region of a black hole spacetime.  These conditions will be 
the same irrespective of whether we are interested in 
asymptotically Lifshitz or adS black holes.  
Assuming that the horizon is non-degenerate, we wish the $g_{tt}$ component 
of the metric to have a simple zero and the $g_{rr}$ to have a simple pole at 
$r=r_+$. Checking that the matter and metric fields and the energy momentum 
tensor are regular at the horizon imposes no further 
constraints and we find the near horizon expansion of the fields to be
\be
\begin{aligned}
F&=f_1(r-r_+)+f_2(r-r_+)^2+...\\
D&=d_1(r-r_+)+d_2(r-r_+)^2+...\\
H(r)&=H_0+H_1(r-r_+)+H_2(r-r_+)^2+...\\ 
\varphi(r)&=\varphi_0+\varphi_1(r-r_+)+\varphi_2(r-r_+)^2+...\\
p(r)&=p_0+p_1(r-r_+)+p_2(r-r_+)^2+...
\end{aligned}
\label{eq:LifshitzBCs}
\ee
where $r_+$ is the Schwarzschild radius of the black hole.  By inserting these 
into the field equations and expanding order by order, appropriate boundary 
conditions can be found.  This procedure leaves us apparently 
with four independent field variables at the horizon: $f_1$, $H_0$, $p_0$ 
and $\varphi_0$ for each choice of $z$ or $\hat m$. 
However, note that $f$ can be shifted by a constant at the price of
rescaling $t$, thus $f_1$ is essentially a gauge degree of freedom,
which is tuned to achieve $F\to1$ at infinity. Note also that the metric 
and field equations are invariant under the rescaling
\be
r\to \lambda r\;\;,\;\;\; t\to\frac{t}{\lambda^z}\;\;,\;\;\; 
x^i\to \frac{x^i}{\lambda}, 
\label{eq:LiRRescaling}
\ee
which means that  we are free to set the Schwarzschild radius of the 
black hole, $r_+ $, to $1$.  We choose, however, to 
keep $r_+$ explicit in our expressions for clarity. 

Numerical solutions can be found by fixing either $\hat m$ (for adS)
or $z$ (for Lifshitz) and using a shooting method to integrate out
from the horizon, tuning the inital data to give a regular asymptotic
solution.

\subsection{AdS Black Holes}

As a warm up, consider first the asymptotically adS solutions. For numerical
simplicity we focus on the case where there is only one unphysical growing
mode at infinity, i.e.\ $\hat m \in[\sqrt{6/5},\sqrt{10}-2]$.  
Note that in this range the exponents of $\varphi$ and $H$ are complex.
Via this process we find a two parameter family of 
asymptotically adS black hole solutions for a fixed value of $\hat m$.  
In light of the previous analytic findings 
these correspond to some combination of gauge and scalar charge.  
A priori we are free to choose any two of  $H_0$, $\varphi_0$ and $p_0$ 
as our free parameters and we shall choose them to be $\varphi_0$ and 
$p_0$. Figures \ref{fig:NumAdSBHHair} to \ref{fig:NumAdSBHChargeAndHair} show 
examples of the asymptotically adS black hole solutions with $\hat m=1.105$. 
\FIGURE{
\centering
\begin{tabular}{cc}
\includegraphics[width=7.25cm]{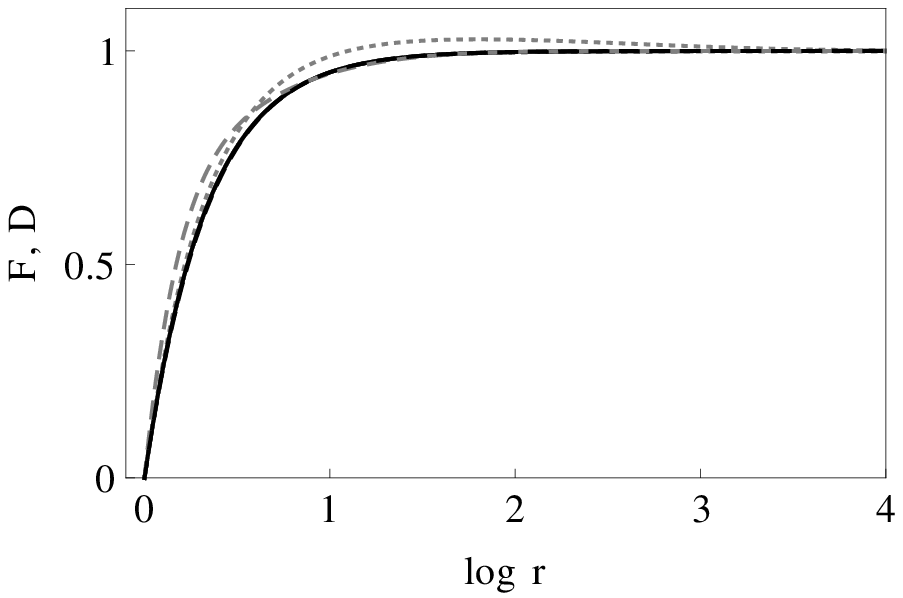}
\includegraphics[width=7.25cm]{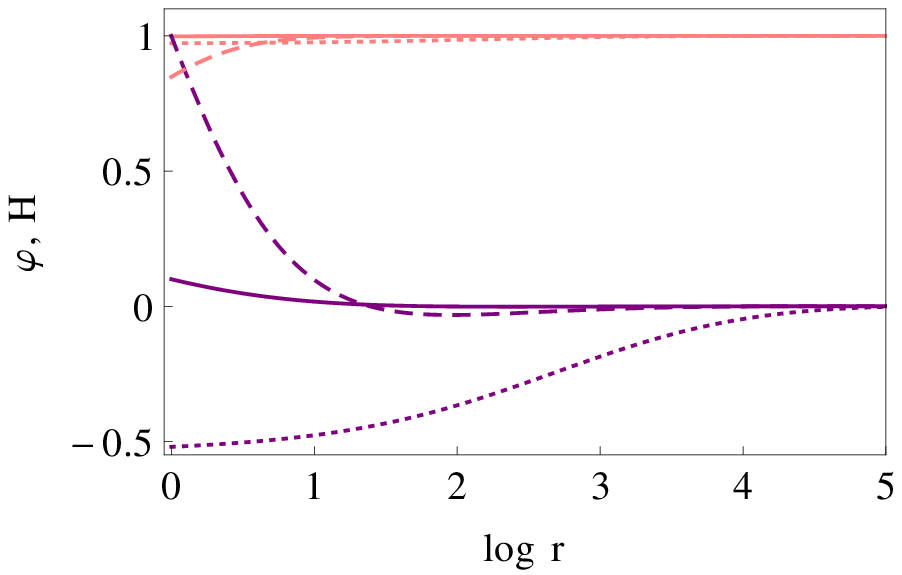}
\end{tabular}
\caption{Field profiles for asymptotically adS black holes 
with $\hat m=1.105$ and no $B-$charge. In the left figure, the black 
lines correspond to the $F$ function and the grey to $D$. On the 
right plot, the purple lines correspond to the dilaton, and the pink to
the $H$ field. In both plots the solid lines are with a small
perturbation, $\varphi_0=0.2$, the dashed and dotted plots to a larger 
dilaton charge, with $\varphi_0=1$ and $\varphi_0=-0.52$ respectively.  }
\label{fig:NumAdSBHHair}
}

Figure \ref{fig:NumAdSBHHair} shows three solutions where $p\equiv0$, in
which the black holes have only scalar charge. One has only a small 
scalar at the horizon, and the other two a more substantial dilaton charge,
one positive and one negative.
The small dilaton perturbation leads to a smaller perturbation in $H$ 
and largely leaves $F$ and $D$ unchanged, which is consistent 
with the approximations made in finding the analytic solution 
(\ref{schypergeo}). Turning $\varphi_0$ up to $1$ shows how positive 
dilaton charge reacts on the geometry. The perturbation in $H$ grows 
and $F$ and $D$ are no longer equal, however, all three functions remain
monotonically increasing outside the horizon. 
Something more interesting happens however if we try to lower the 
value of the scalar at the horizon, i.e.\ letting $\varphi_0<0$. 
Overall, very little happens to the geometry, however, as can be seen
from the plot for $\varphi_0=-0.52$, the $D$ and $H$ functions cease to
be monotonic and all the fields relax to their asymptotic values
significantly more slowly than for positive scalar charge.
Indeed, there is a critical value of $\varphi_0 \simeq -0.53$, below
which the charged black hole solution ceases to exist. 
This is because there is a runaway behaviour in ${\hat V}(\varphi)$ 
for $\varphi$ too negative. This critical value shifts towards the 
origin as we turn on $B-$charge, as can be seen by looking at the 
source for the $\varphi$ equation of motion at the horizon.

Figure \ref{fig:NumAdSBHCharge} shows solutions with $B-$charge, while
keeping $\varphi_0=0$. Again, we show the comparison between a small 
and larger gauge charge. The small perturbation in $P$ leaves the 
other fields largely unchanged, and comparing the analytic approximate
and numerically generated profiles of $P$ for the same value of $\hat m$, 
one finds that the two appear identical to the naked eye.  
Increasing $p_0$ alters all the other fields as seen in the dashed
plots, and once again $F$ and $D$ increasingly differ. 
\FIGURE{
\centering
\begin{tabular}{cc}
\includegraphics[width=7.25cm]{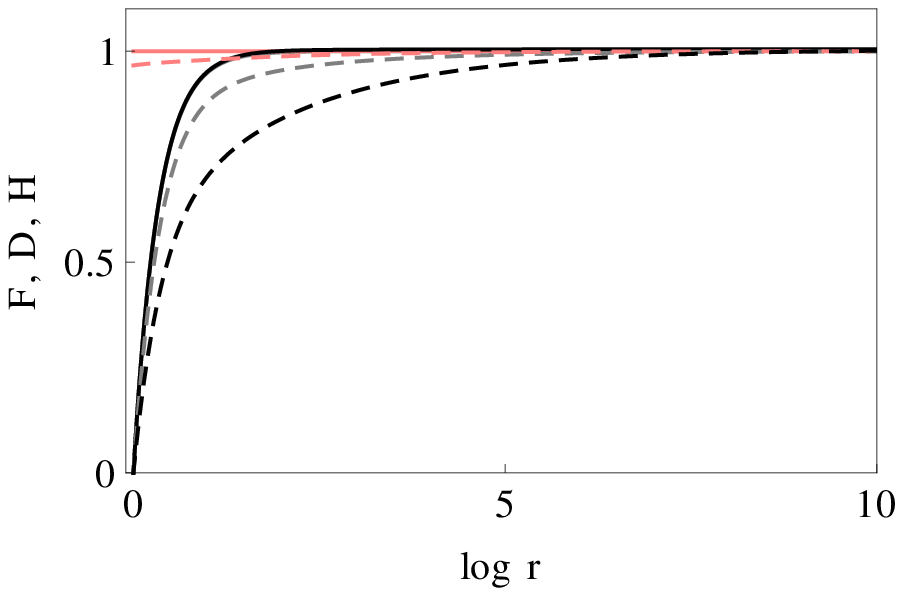}
\includegraphics[width=7.25cm]{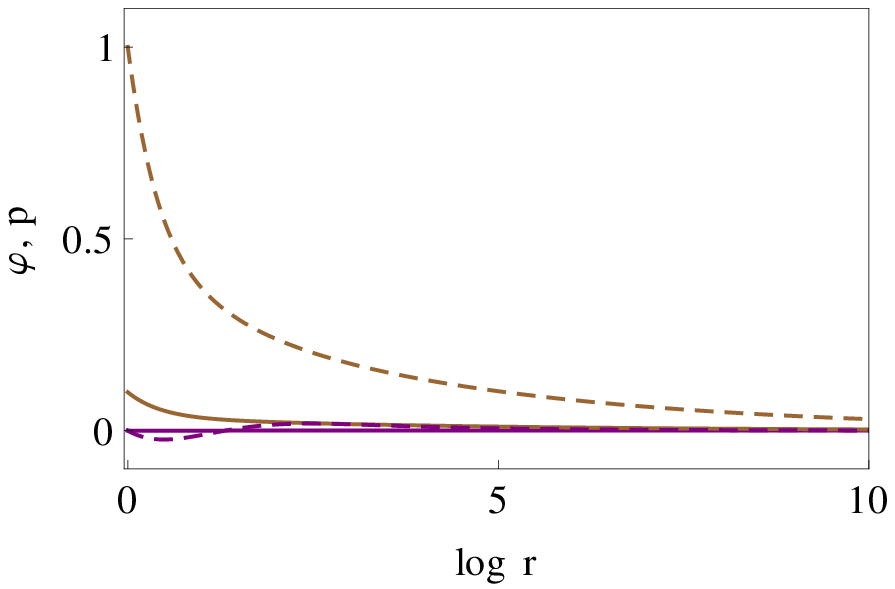}
\end{tabular}
\caption{Field profiles for asymptotically adS black holes 
with $\hat m=1.105$, with $\varphi_0=0$.  As before, the black, grey,
pink and purple lines correspond to the $F$, $D$, $H$ and $\varphi$ fields, 
with the $p$-field being plotted in brown. The solid lines are for a 
small perturbation, $p_0=0.1$, and the dashed to a larger charge, $p_0=1$.}
 \label{fig:NumAdSBHCharge}
}

The effect of both scalar and gauge charge on the black hole is shown in 
figure \ref{fig:NumAdSBHChargeAndHair}. Here, the dotted line shows
a nearly critical negative scalar charge black hole, in which the $B-$charge
has been turned to near extremality (the temperature of this black hole
is $0.024 r_+$). 
\FIGURE{
\includegraphics[width=7.25cm]{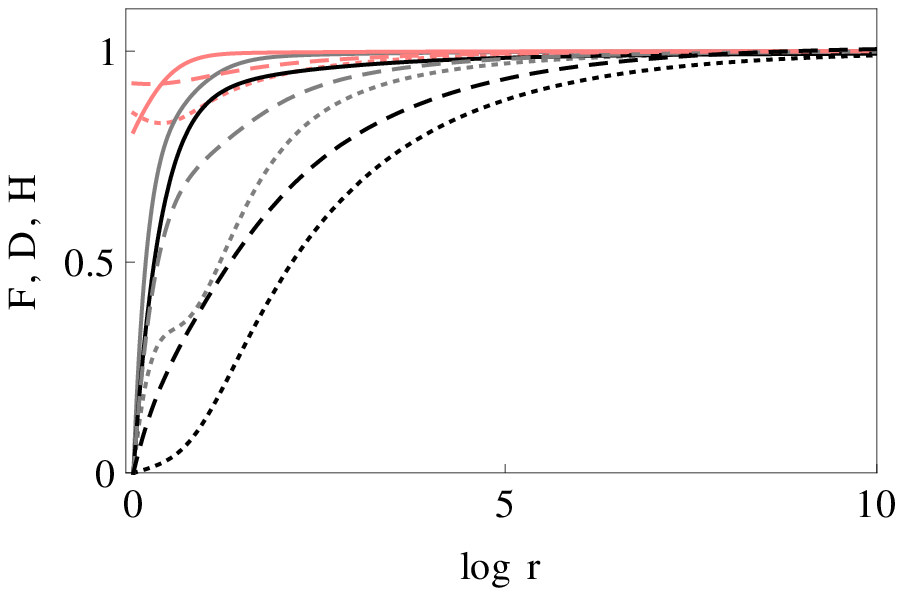}
\includegraphics[width=7.25cm]{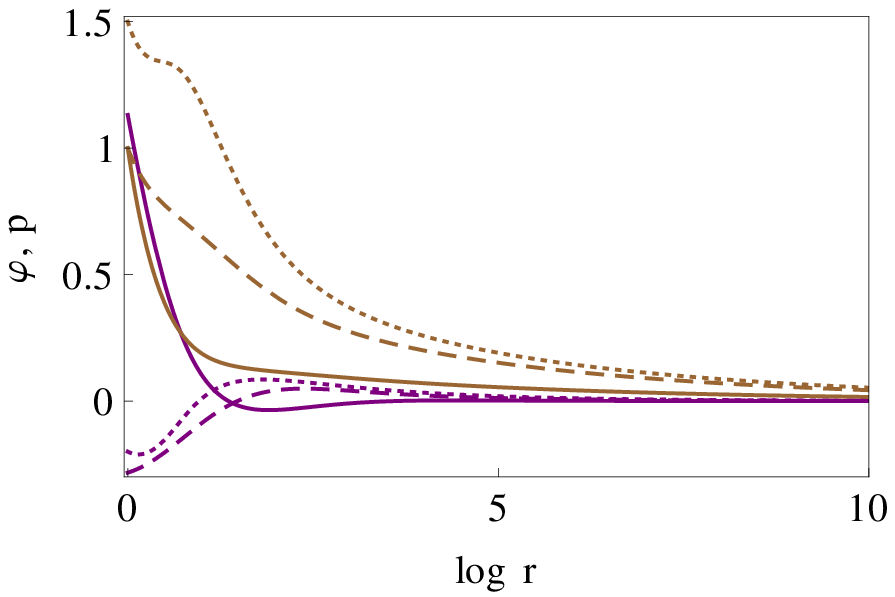}
\caption{Field profiles for generic adS black holes with $\hat m=1.105$.
The solid lines correspond to horizon data $\varphi_0=4\sqrt{2}/5$, 
$p_0=1$; the dashed lines to $\varphi_0=-\sqrt{2}/5$, $p_0=1$; and
the dotted lines to $\varphi_0=-0.2$, $p_0=1.5$. The colour coding
of the plot is the same as figure \ref{fig:NumAdSBHCharge}.}
\label{fig:NumAdSBHChargeAndHair}
}

\subsection{Lifshitz Black Holes}

We  now turn to black hole solutions that asymptote 
the Lifshitz spacetime defined by (\ref{eq:lifsolns}).  A crucial difference 
between this case and the asymptotically adS case is that the background
2-form gauge field is now nonzero, $p\neq0$.  
As a result we were unable to find an exact analytic expression for the 
black hole similar to (\ref{eq:adssch}), only perturbative or asymptotic
solutions. We must therefore rely solely on numerical results for 
full solutions to the equations of motion. Another difference with the adS
case is that all of the fields necessarily participate in the black hole
solution. That this is the case can be seen by checking the eigenvectors
of the perturbations around the critical point. All of the eigenvectors have
(different) combinations of scalar, gauge and geometry components. Thus
we expect that, for Lifshitz black holes, nontrivial profiles of {\it all} 
of the fields will generally be present.

The numerical solutions are found in precisely the same way as 
in the adS case with the only difference being that the parameters 
of the theory are now defined by (\ref{eq:lifsolns}) where the 
dynamical exponent $z\geq1$ is used to fix the theory as opposed 
to $\hat m$.  For simplicity we present solutions corresponding to 
the upper sign choice in (\ref{eq:lifsolns}) as this joins the adS 
branch and is more stable to integrate. 
By integrating the equations of motion (\ref{eq:phi}) to 
(\ref{eq:P}) we once again find a two parameter family of asymptotically 
Lifshitz black hole solutions for each value of $z$ to which we assign the free 
parameters $\varphi_0$ and $p_0$.  Using intuition from the adS case we 
suggest that these parameters relate to the scalar and $B-$charge 
of the black hole, however, since all the fields participate in any
asymptotic fall-off to the Lifshitz spacetime, this relation will
not be completely straightfoward. 
\FIGURE{
\begin{tabular}{cc}
\includegraphics[width=7.25cm]{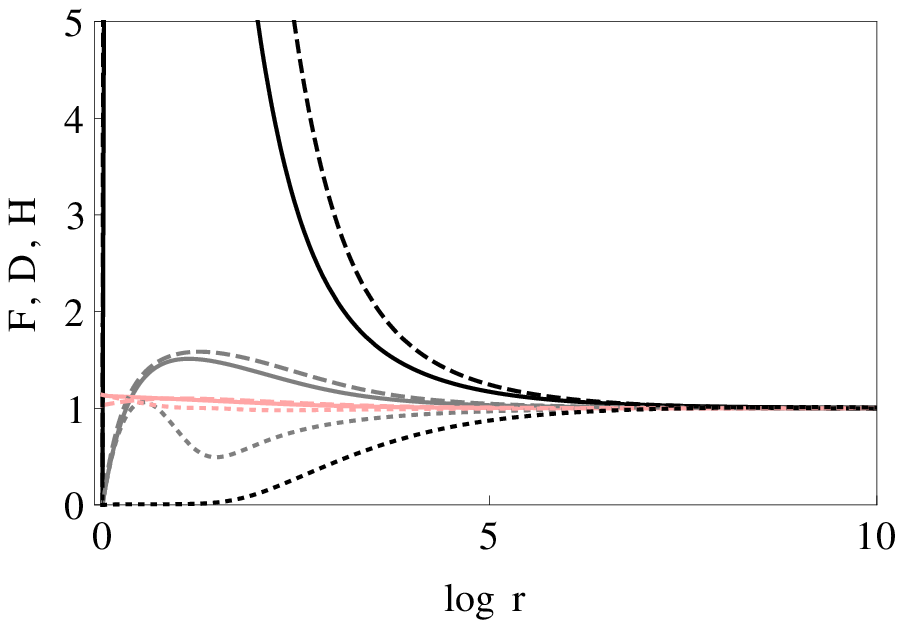}
\includegraphics[width=7.25cm]{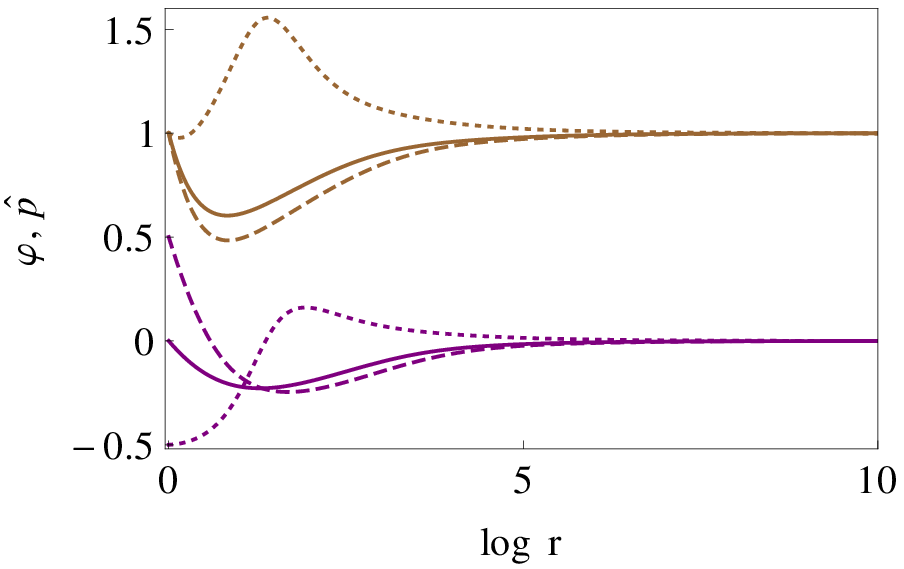}\\
\includegraphics[width=7.25cm]{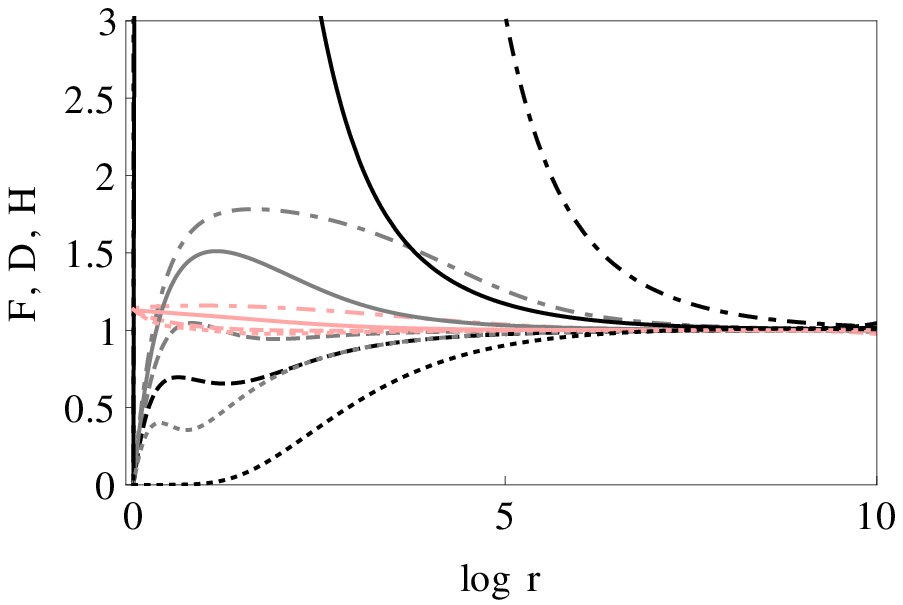}
\includegraphics[width=7.25cm]{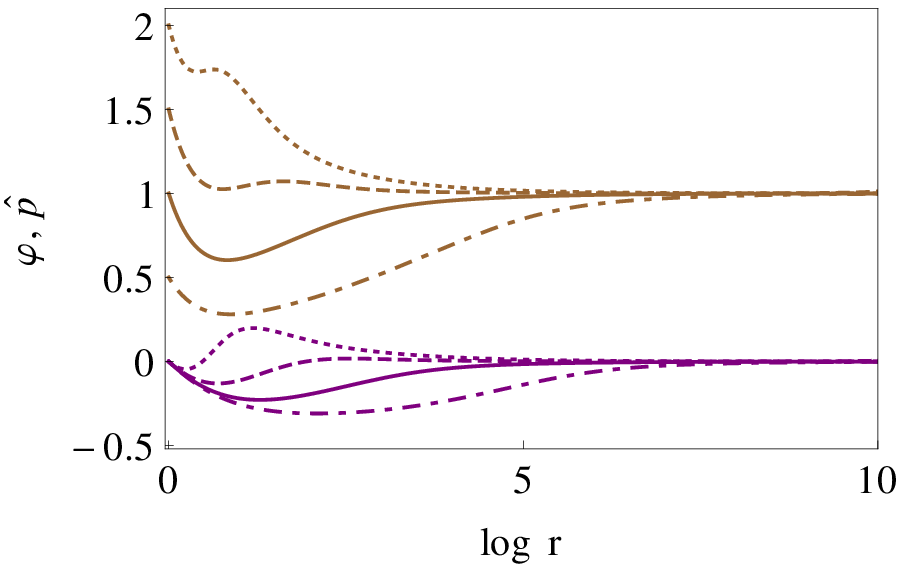}\\
\end{tabular}
\caption{A selection of plots showing the field profiles for 
asymptotically Lifshitz black holes for $z=2$. The colours are
the same as for ads, with $F$, $D$, $H$, $\varphi$ and ${\hat p} =
p/{\hat b}$ being displayed in black, grey, pink, purple and brown
respectively. The upper pair of plots explore the effect of varying
$\varphi_0$, and the lower plots the impact of changing ${\hat p}_0$.
In each case, the labelling of the curves is defined by the initial
conditions in the right hand plot.}
\label{fig:NumLifBHz=2}
}

In all of the Lifshitz plots, we renormalize the $z$-dependence
of the gauge field by plotting ${\hat p} = p/{\hat b}$, so that
unnecessary variation with $z$ is scaled out.
Figure \ref{fig:NumLifBHz=2} explores the impact of varying the
gauge and scalar initial conditions on a $z=2$ Lifshitz black hole.
The plots are reasonably self-explanatory, exploring the impact of 
altering $\varphi_0$ (upper) and $\hat{p}_0$ (lower) relative to
the fiducial black hole solution shown in each case by solid lines,
whose horizon values of the dilaton
and $B-$field are the same as the asymptotic values. 
Most of these black hole solutions (see also figure \ref{fig:NumLifz})
have extremely strongly warped geometries near the horizon, with the 
``Newtonian potential", $F$, rising very sharply to a rather high
maximum before falling to its asymptotic value of $1$ from above.  
For example, in the fiducial solution, the maximum of $F$ is 
around $30$ (in units of $r_+$), and lowering the horizon value of 
$\hat{p}_0$ only exacerbates this effect.

In the upper plots, showing some sample solutions for different $\varphi_0$, 
we see that decreasing $\varphi_0$ below zero damps the $F$ potential,
which now looks more like a canonical black hole, with the example
shown in the dotted plot, $\varphi_0 = -0.5$ being representative of
a nearly `extremal' black hole, in the sense that the solution will
cease to exist if $\varphi$ is lowered further (for the same reason as
in the adS case) and also in the sense of the temperature dropping to zero.
Increasing $\varphi_0$ on the other hand has the opposite effect, with
$\varphi_0=0.5$ increases the sharp peak of $F$, which now has a 
maximum of around $80$ (the dashed lines). 
Interestingly however, this variation with the dilaton horizon value is
not monotonic, and as $\varphi_0$ is increased further, the amount
of warping peaks, then subsides, and as we will see in the
next section when we consider thermodynamics, this behaviour is
mirrored in the temperature of the black hole dropping to zero.

In the lower pair of plots, which explore changing the horizon value 
of the gauge field, we see that increasing $\hat{p}_0$ rapidly restores 
the $F$ field to a more canonical form,
in the case of the dashed (${\hat p}_0 = 1.5$) and dotted (${\hat p}_0=2$)
plots. Since the temperature of these black holes drops, increasing 
the horizon value of ${\hat p}$ can be seen to be analogous to charging
up a Reissner-Nordstrom black hole. Correspondingly, as can be seen
in the dot-dashed plot, dropping ${\hat p}_0$ below its asymptotic 
value causes the black hole to become more strongly warped, and as 
we will see, hotter. 

Altering both $\varphi_0$ and ${\hat p}_0$ produces a combination
of these effects: increasing $\varphi_0$ first moves the black hole away
from, then towards, `extremality', and decreasing ${\hat p}_0$ always moves
the black hole away from extremality. Correspondingly, 
the maximal value of $B-$charge (tracked by ${\hat p}_0$) first increases,
then decreases, as we increase the dilaton charge $\varphi_0$.
\FIGURE{
\centering
\includegraphics[width=7.25cm]{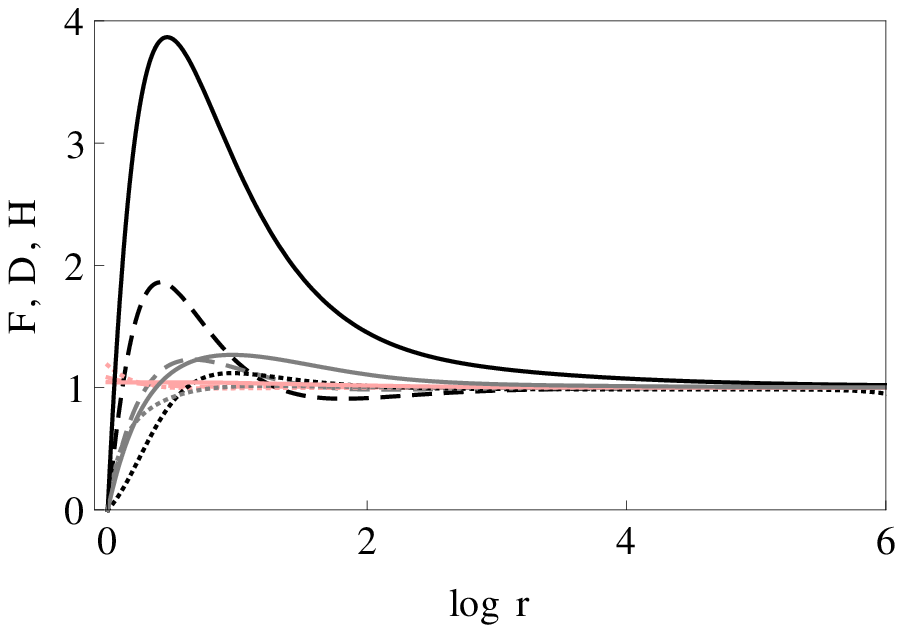}
\includegraphics[width=7.25cm]{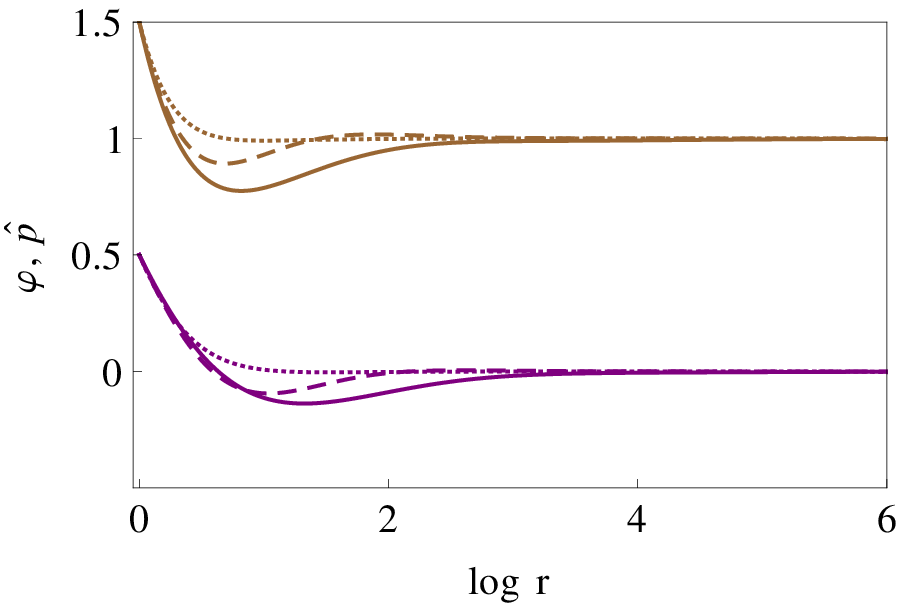}
\caption{Plots showing the field profiles for asymptotically Lifshitz 
black holes with $\varphi_0=0.5$, $\hat{p}_0=1.5$ at different $z$.
The black, grey, pink, blue and red lines correspond to $F$, 
$D$, $H$, $\hat{p}$ and $\varphi$ respectively.  The solid lines 
correspond to $z=2$, the dashed lines to $z=3$, and the 
dotted lines to $z=5.75$. }
\label{fig:NumLifz}
}

Finally, the plots in figure \ref{fig:NumLifz} show the effect of altering the 
dynamical exponent, $z$, for a system with $\varphi_0=0.5$ 
and $\hat{p}_0=1.5$.  The plots show the field profiles for $z=2$, $z=3$ 
and $z=5.75$, where the last value was chosen to be within the range 
for which all the eigenvalues of figure 
\ref{fig:liffp} are real. These plots show how $z$ can alter 
the field profiles, particularly $F$ and $D$,  and that 
increasing $z$ hastens the convergence to the Lifshitz solution. 
This was to be expected since, as can be seen 
in figure \ref{fig:liffp}, increasing $z$ largely reduces the 
eigenvalues governing each fields approach to the Lifshitz fixed point.

\subsection{Uplifting to Type IIA in 10 dimensions}

With the 6D solutions in hand, it is straightforward to 
uplift them to configurations in Type IIA massive supergravity.  
Following \cite{IIAred,GPTZ}, we define
\bea 
&& X(r) = e^{\phi_0/\sqrt{2}}\left(\frac{g}{3\m}\right)^{1/4}
e^{\varphi(r)/2}  \\
&& \texttt{C} (\rho)= \cos \rho \,, \qquad \texttt{S}(\rho)= \sin \rho \\
&& \Delta(\rho) = X \, \texttt{C}^2 + X^{-3}\,\texttt{S}^2  \\ 
&&  U(\rho) = X^{-6}\,\texttt{S}^2 - 3 X^2\,\texttt{C}^2 
+ 4 X^{-2}\,\texttt{C}^2 -6 X^{-2}  \, ,
\eea
as well as the constant $k=\left(3\m g^3\right)^{1/4}/2$.  We can then write
the ten dimensional, uplifted configurations as:
\bea
&& ds_{10}^2  = \texttt{S}^{1/12} \,X^{1/8}\left[  
\Delta^{3/8}(LiBH_4\times {\Omega}_2) - 2 k^{-2} \Delta^{3/8} X^2\,d\rho ^2 
- \frac12 k^{-2} \Delta^{-5/8} X^{-1} \texttt{C}^2
\sum_i^3 (h^{(i)})^2 \right] \, , \nonumber \\
&& {\bf F_4} = \frac{\sqrt{2}}{6}\,k^{-3} \, \texttt{S}^{1/3} 
\, \texttt{C}^3  \, \Delta^{-2} \,U \, d\rho  \wedge  \epsilon_3  
+ \sqrt{2} \,k^{-1}\,\texttt{S}^{1/3}\texttt{C} \, 
X^4\,\star_6 G_3 \wedge d\rho\nonumber \\ 
&&\qquad- \frac{1}{\sqrt{2}} \,k^{-2}\, \texttt{S}^{1/3} \,  
\texttt{C} \, F_2^{(3)} \wedge h^{(3)}\wedge d\rho
+ \frac{1}{4\sqrt{2}} \,k^{-2}\, \texttt{S}^{4/3} \, 
\texttt{C}^2 \Delta^{-1} \,X^{-3}\,F_2^{(3)} \wedge
\sigma^{(1)}\wedge \sigma^{(2)} \nonumber \\
&&\qquad + \sqrt{2}\, k^{-3}\, \texttt{S}^{4/3} \texttt{C}^4 
\Delta^{-2} X^{-3} dX \wedge \epsilon_3 \, ,\label{ulift1} \\
&& {\bf G_3} = 2 \sqrt{2} \frac{k^2}{g}\, \texttt{S}^{2/3}  \, 
G_3  \, , \quad {\bf F_2} = 0 \, ,
\nonumber \\
&&  e^{\Phi} = \texttt{S}^{-5/6} \,  \Delta^{1/4} \,  X^{-5/4} \,,\nonumber
\eea
where 
\be
h^{(i)}  = \sigma^{(i)} - g \, A_1^{(i)}\,,
\ee
with $\sigma^{(i)} $ the left-invariant 1-forms on $S^3$, 
and $\epsilon_3 =h^{(1)}\wedge h^{(2)} \wedge h^{(3)}$.   
The parameters of the 6D theory are related to the Type
IIA mass parameter via ${\bf m} = 
\left(2 \,{\m} \, g^3/27\right)^{1/4}$.  The uplift gives us some 
insight into the kinds of sources in 10D that give rise to the adS 
and Lifshitz black holes in 4D.
Notice that the ten dimensional RR ${\bf F_2}$ field strength vanishes,
while the  RR  ${\bf F_{4}}$ field and the NS ${\bf G_3}$ field are
switched on in several directions.  The main difference between 
the black hole solutions discussed here and their pure Lifshitz/adS 
counterparts presented in \cite{GPTZ} is in the 
non-trivial $r$-profiles, which imply that ${\bf F_{4}}$ 
contains an additional component in  the direction $dX\wedge\epsilon_3$. 
The configurations could be interpreted as a system of D-branes 
and NS-branes, analogously to \cite{GPTZ}.

\section{Thermodynamics}

Having determined the black hole solutions for our supergravity set-up, 
it is interesting to investigate some of 
their general properties besides their field profiles.  
In this section we make the first 
few steps of this exploration by studying the dependence of the 
temperature of these black holes on the initial parameters $\varphi_0$ 
and $p_0$, as well as commenting on entropy.

The temperature of the black hole is given by
\begin{align}
T= \frac{r_+^{z+1}}{4\pi}\sqrt{D'F'} \;|_{\,r=r_+}\,,
\end{align}
which is calculated directly from the numerical solutions.
Figures \ref{fig:TempAdsBH} and \ref{fig:TempLifBH} show the
temperatures of both asymptotically adS, and asymptotically
Lifshitz black holes as a function of our initial parameter ${\hat p}_0$
and $\varphi_0$. In all the temperature plots, we have shown the 
temperature normalized at $r_+=1$ for simplicity. We discuss the 
$r_+$ dependence of the temperature at the end of this section.
\FIGURE{
\includegraphics[width=7.25cm]{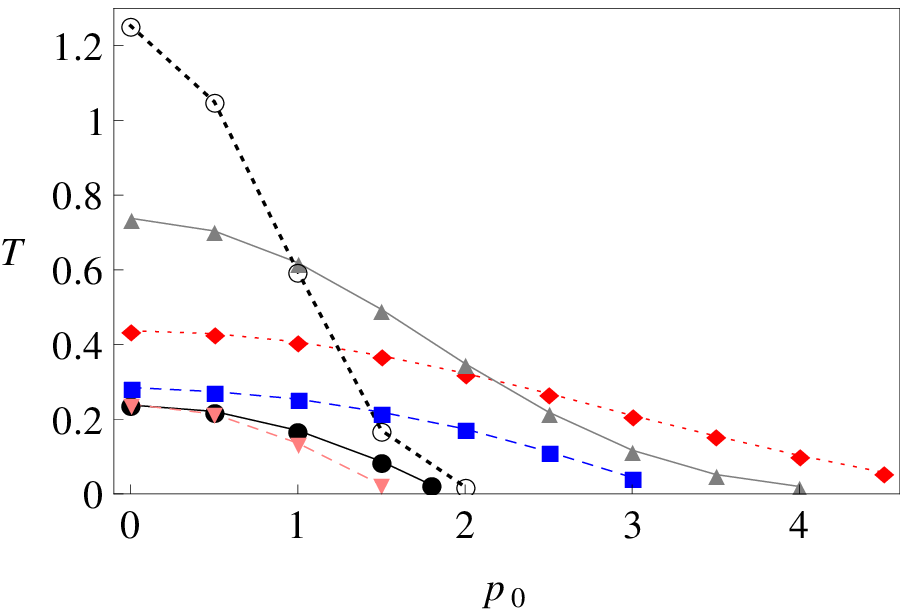}
\includegraphics[width=7.25cm]{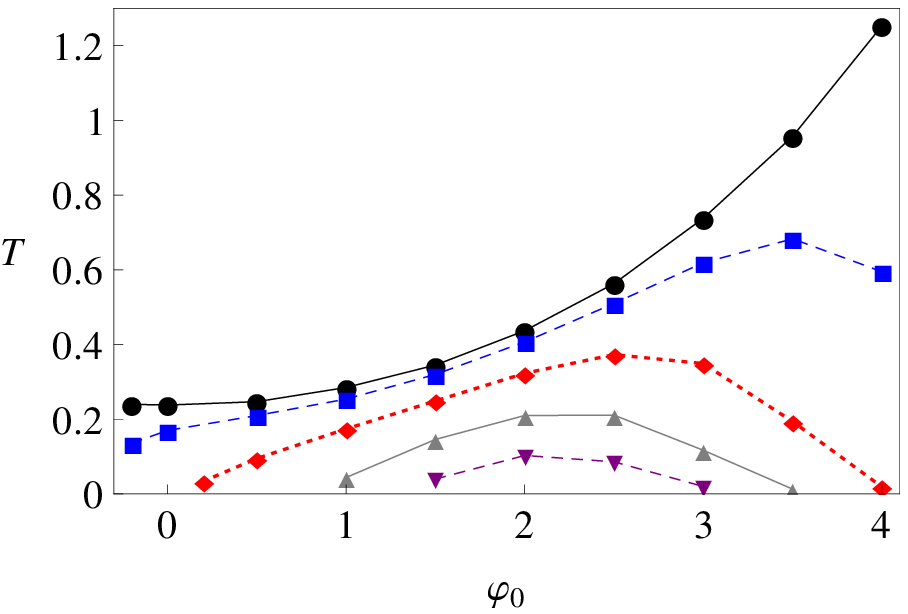}
\caption{Plots of the temperature of an adS black hole with $\hat 
m=1.105$ shown as a function of $p_0$ (left) and $\varphi_0$ (right). 
In the left plot, the dashed pink data  in inverted triangles
corresponds to $\varphi_0 = -0.2$; the black data in dots to $\varphi_0=0$; 
the dashed blue with squares to $\varphi_0=1$; the dotted red 
with diamonds to $\varphi_0=2$; the grey triangles to $\varphi_0=3$;
and the dashed black with open circles to $\varphi_0=4$. 
In the right plot, the lines run from ${p}_0=0$
in black with dots at the top, to ${p}_0=4$ in dashed purple with 
inverted triangles at the bottom in increments $\Delta{p}_0=1$.}
\label{fig:TempAdsBH}
}

In figure  \ref{fig:TempAdsBH}, plots are shown of the temperature of
an asymptotically adS black hole with $\hat m=1.105$. On the left, 
the plot is shown as a function of the $B-$charge for a range of 
$\varphi_0$, on on the right as a function of $\varphi_0$ for a range
of $p_0$. The left plot shows the expected behaviour of a charged
black hole, in that adding charge reduces temperature monotonically to
zero at an extremal limit. The effect of scalar charge in this case
is more interesting. At zero $B-$charge, the impact of increasing
$\varphi_0$ is to increase the temperature of the black hole, and
one might expect therefore that the allowed $B-$charge range is 
increased. However, an interesting phenomenon occurs. As the black 
hole becomes more and more charged under the scalar field, the maximal
amount of $B-$charge we are able to add starts to drop, 
and at very high scalar charges we can no longer add much gauge charge.
That the ``extremal'' limit should not be a simple sum of the two
charges, but some more complex combination is an interesting difference
from most black holes with more than one charge.

Figure \ref{fig:TempLifBH} shows the corresponding plots
for a Lifshitz black hole, with $z=2$ taken as an example.
As with the adS black hole, increasing ${\hat p}_0$ reduces the temperature,
and once again, we see that there is a finite
range of  $\varphi_0$ for which the black holes exist. 
However, because the Lifshitz spacetime has a nonzero background
$B-$field, there is a clear difference in the temperature as a
function of ${\hat p}_0\to0$. For the Lifshitz black hole, the 
temperature increases sharply as we reduce the initial value 
of ${\hat p}_0$, and would appear to diverge as ${\hat p}\to 0$. 
In the absence of analytic arguments we cannot say definitively that
$T$ diverges, however, our numerical integrations become more and
more extreme as we reduce $\hat{p}_0$. It is worth noting that we
have presented our temperature plots renormalized to $r_+=1$, clearly,
dropping $r_+$ drops the temperature, so it is possible that we can
achieve $\hat{p}_0 \to0$ by taking $r_+ \to 0$, indeed, such a
spacetime would represent a flow from a Lifshitz space in the UV
to an adS space in the IR,  \cite{BGR}. However, the results of
\cite{BGR} would indicate that this will only happen for a specific
value of $\varphi_0$, namely, the one corresponding to an adS solution.
Thus, we would expect the generic $\hat{p}_0\to0$ limit to be
singular.
\FIGURE{
\includegraphics[width=7.25cm]{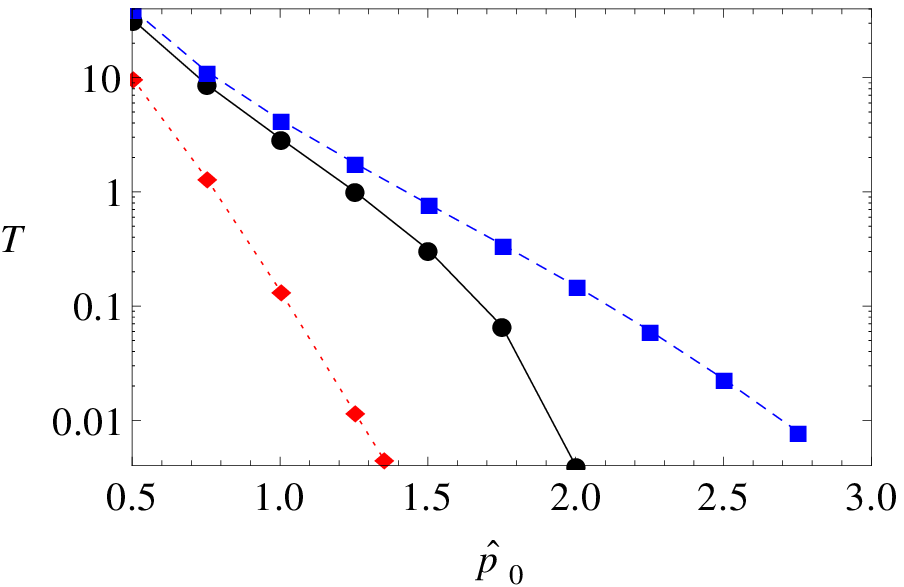}
\includegraphics[width=7.25cm]{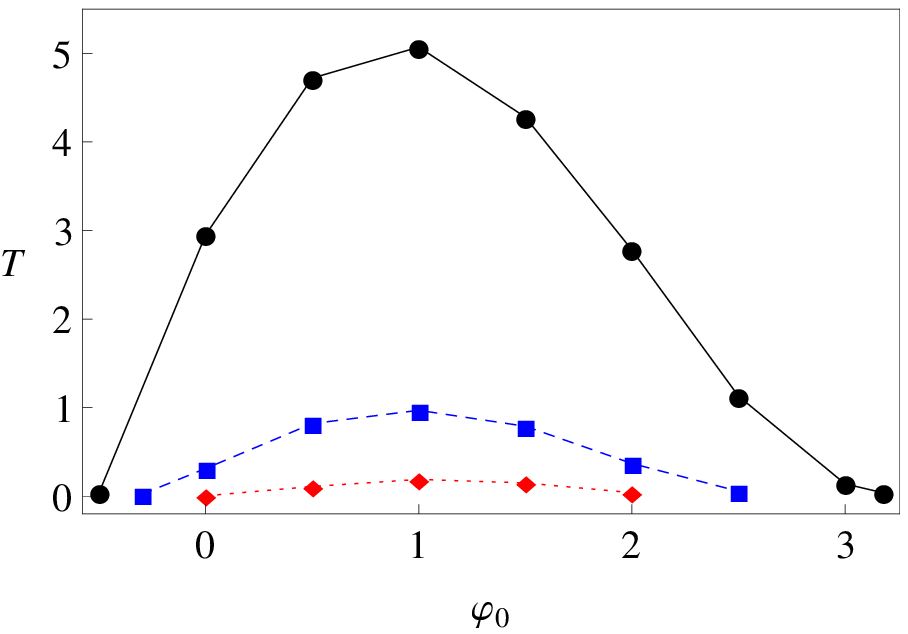}
\caption{Plot of the temperature of asymptotically Lifshitz black holes 
for varying ${\hat p}_0$ and $\varphi_0$, with $z=2$. On the 
left the plot depicts the temperatures of $z=2$ black holes as a function
of ${\hat p}_0$ for $\varphi_0=0$ in black with circular data points,
$\varphi_0=1.5$ in dashed blue with square data points, and 
for $\varphi_0=3$ in dotted red with diamond data points.
On the right, the temperature is shown as a function of $\varphi_0$ 
for ${\hat p}_0=1$ in black (circles), ${\hat p}_0=1.5$ in blue 
(dashed/squares), and ${\hat p}_0=2$ in red (dotted/diamonds).  }
\label{fig:TempLifBH}
}

In figure \ref{fig:multiZ} we see how altering $z$ affects the 
temperature of the black hole by looking at the variation of
temperature with ${\hat p}_0$ for sample values of $z$, and 
exploring in detail the $z$-dependence for sets of representative
initial data. In general, we see that increasing $z$ raises 
the temperature for small 
$\hat p_0$ but lowers it for larger $\hat p_0$, although
for temperatures close to $1$, the temperature seems to first increase
then decrease with $\hat{p}_0$. 
\FIGURE{
\includegraphics[width=7.25cm]{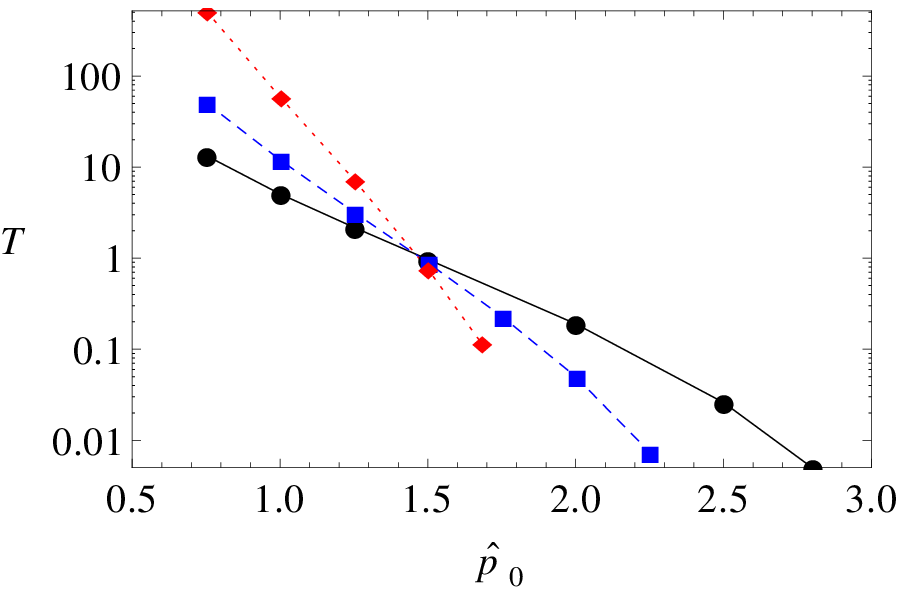}
\includegraphics[width=7.25cm]{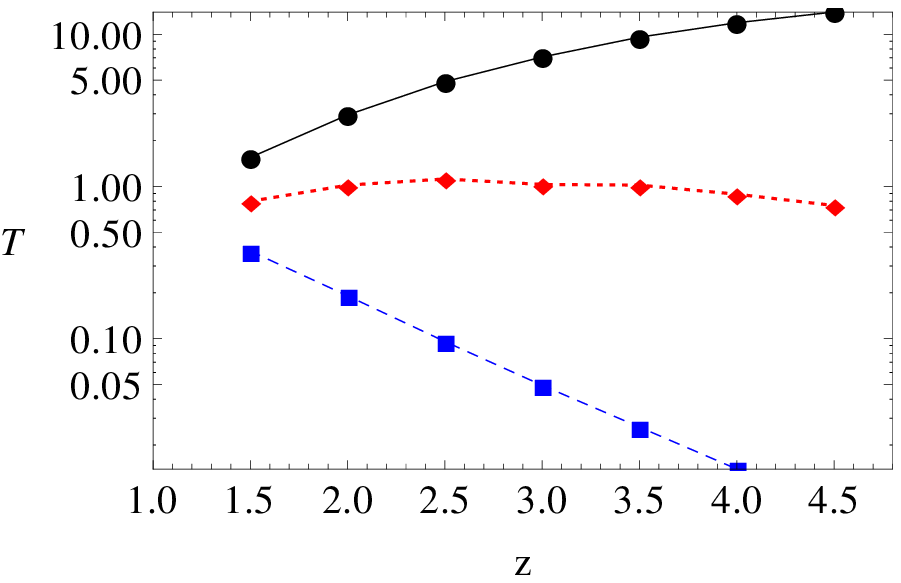}
\caption{Plot of the temperature of asymptotically Lifshitz black holes 
for varying ${\hat p}_0$, $\varphi_0$, and $z$. 
On the left, the curves give the temperature of a Lifshitz black
hole with $\varphi_0=1$ as a function of ${\hat p}_0$ for 
different values of $z$: black (circles) to $z=2$, blue (dashed/squares) 
to $z=3$, and red (dotted/diamonds) to $z=5$.  The three curves appear to 
intersect at a single point, however, the resolution of the data is
insufficient to be sure if this is exact.
On the right, the temperature is plotted
as a function of $z$ for initial data $(\varphi_0,\hat{p}_0)=(0,1)$
in black (circles), $(\varphi_0,\hat{p}_0)=(1,2)$ in blue (dashed/squares),
and $(\varphi_0,\hat{p}_0)=(0,1.25)$ in red (dotted/diamonds).}
\label{fig:multiZ}
}

Finally, we consider the entropy of the black holes, which 
can be computed from the area of the horizon as:
\be
S = \frac14 r_+^2 H (r_+)
\ee
per unit volume (and setting $G_6=1$). 
Figure \ref{fig:SvT} shows how the entropy of a $z=2$ Lifshitz
black hole varies as a function of temperature at fixed $r_+=1$
for different values of initial data. The two plots show set of
data for $S(T)$ having fixed $\varphi_0$ and varying $\hat{p}_0$ 
(left) or data for fixed $\hat{p}_0$ and varying $\varphi_0$
(right). In the left plot, we see that just
as varying $\hat{p}_0$ at fixed $\varphi_0$ has a much more 
uniform effect on temperature, so varying $\hat{p}_0$ at fixed $\varphi_0$
has a somewhat more consistent effect on entropy, although curiously
the entropy generally drops as we increase the temperature. For
$\varphi_0=0$ however, the entropy remains fairly constant.
Since we have fixed $r_+=1$, we would not necessarily expect the
entropy to vary hugely with $\hat{p}_0$, at least by analogy with the 
Reissner-Nordstrom solution.

When exploring the $S(T)$ plot for varying $\varphi_0$ however, the picture 
becomes much more interesting. We would expect entropy to vary much
more strongly with $\varphi_0$ at fixed $r_+$, since we have seen
from our eigenvalue analysis of the perturbations that the two ``scalar''
modes, the dilaton and the internal breather $H$, are very much
coupled by the equations of motion. We therefore expect that
altering $\varphi_0$ will alter $H(r_+)$ and hence the entropy
to a much greater extent, and this is indeed what we see.
However, what is interesting is the modulating behaviour of both the
temperature and entropy as a function of $\varphi_0$. 
We see that at a given temperature and $\hat{p}_0$, there are
two possible values for the scalar charge, one with higher entropy
that the other. Although it is not entirely clear from the plot, the
curves have $\varphi_0$ increasing in a clockwise direction, hence it
is the black hole with lower $\varphi_0$ that is entropically preferred.
This indicates that these black holes will likely have scalar
instabilities, perhaps shedding scalar charge to increase their overall
horizon area. How this is consistent with the usual concept of a
black hole accreting to increase its area might prove an interesting
investigation.

Finally, we should comment on the impact of varying $r_+$:
Because of the scaling symmetry present in the equations of motion, 
all of the numerically computed fields are dependent
on $r/r_+$, and thus simply stretch with $r_+$. In particular,
the horizon value of $H$ does not change with $r_+$, and the
derivatives of $F$ and $D$ just scale as $1/r_+$. Thus, the
entropy scales as $r_+^2$, independent of the value of $z$, and
the temperature as $r_+^z$. (Of course, the entropy and temperature
vary with the initial data of the charges as we have seen.)
The variation of entropy with temperature is therefore explicitly
the expected relation $S \propto T^{2/z}$ for a field theory in 
flat 2+1 dimensions.
\FIGURE{
\includegraphics[width=7.25cm]{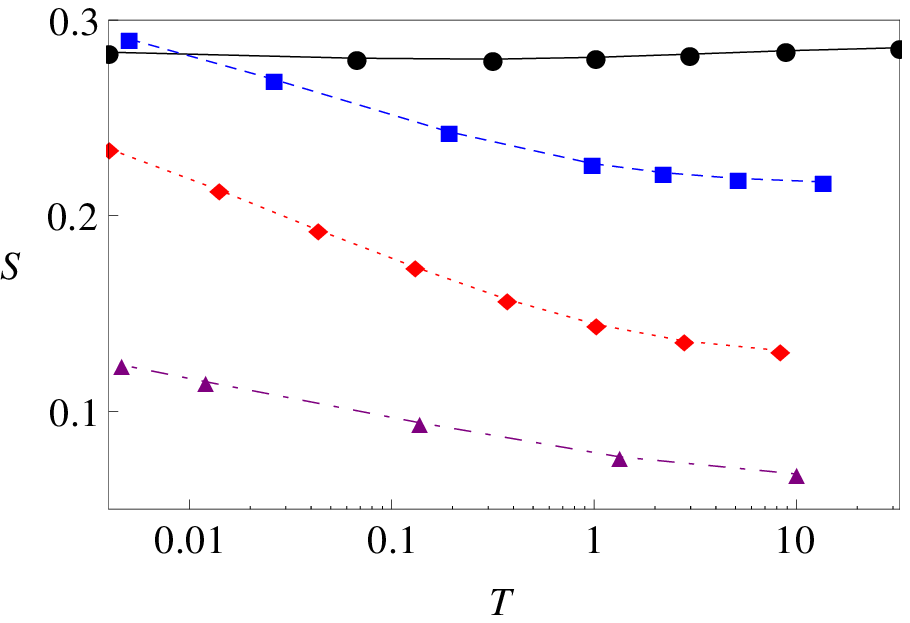}
\includegraphics[width=7.25cm]{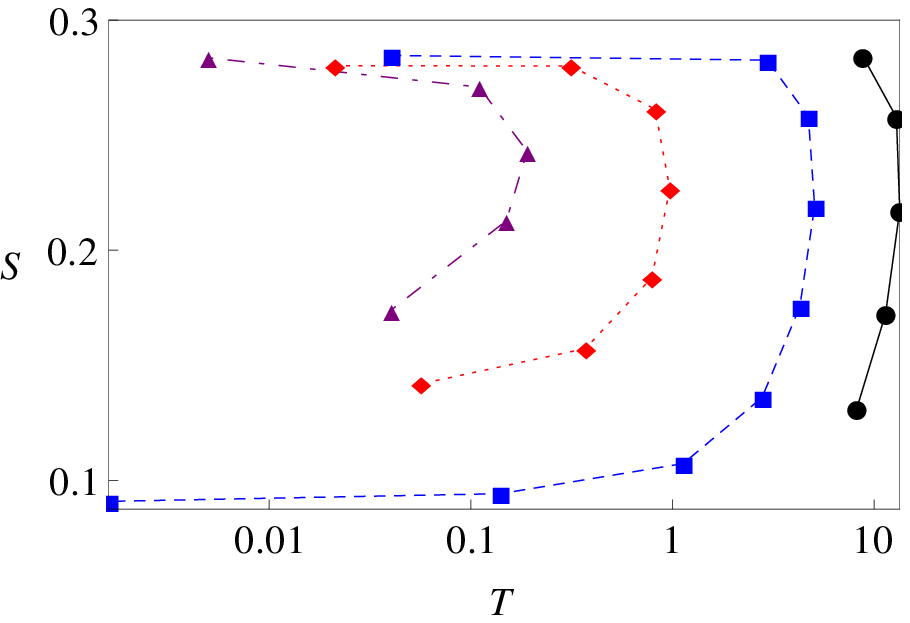}
\caption{Plots of the entropy of a $z=2$ Lifshitz black hole as a
function of the temperature with $r_+=1$. On the left, the data points
represent different values of ${\hat p}_0$ for $\varphi_0 = 0,1,2$, and $3$,
in black (dots), dashed blue (squares), dotted red (diamonds) and 
dot/dash purple (triangles) respectively. On the right, the data points 
represent different values of $\varphi_0$ for $\hat{p}_0 = 0.75,1,1.5$ 
and $2$, in black (dots), dashed blue (squares), dotted red (diamonds) and 
dot/dash purple (triangles) respectively. 
}
\label{fig:SvT}
}

\section{Conclusions}

In this paper we have built string  black hole configurations
that asymptote Lifshitz geometries for general dynamical exponents $z>1$.
We began by finding these solutions in a supergravity theory that 
corresponds to a consistent truncation of massive type IIA 
string theory, namely the maximal gauged six dimensional supergravity 
of Romans, and then uplifted them to massive type IIA supergravity.  
They are related to asymptotically adS black hole solutions
in the same supergravity, also studied here, and extensions of 
the pure Lifshitz geometries found in \cite{GPTZ}.  

Our supergravity theory has rich field content, and yet allows simple
pure adS$_4 \times H_2$ and Li$_4 \times H_2$ compactifications.   
By expressing the full set of equations in 
the form of an autonomous dynamical system, we studied 
perturbations around these exact solutions, determining  
analytically all the possible asymptotic behaviours of static, 
planar adS and Lifshitz black holes for our theory. 
We found that various asymptotics are allowed, depending on which fields 
are switched on and on the values of the parameters involved, in 
particular the dynamical exponent $z$. 

Starting with the adS case, one exact analytic solution is of 
course known: adS-Schwarzschild.  By studying this background, 
and exciting the supergravity two-form potential $B_2$ and 
dilaton in the probe limit, we acquired insight into the 
properties of adS black holes. In particular, apart from the 
horizon size, the black holes are characterized by two quantities, 
$\varphi_0$ and $p_0$, 
which we are motivated to call scalar and $B$-charge.  
After exploring the adS case, we applied the intuition gained 
from it to analyze Lifshitz configurations, for arbitrary
dynamical exponent $z>1$.
In fact, a non-trivial scalar and $B$-charge is a necessary 
ingredient for obtaining asymptotically Lifshitz geometries.  
Although the notion of a charged 
black hole in the Lifshitz case is less straightforward than for adS, 
since any Lifshitz configuration is always characterized by the presence
of all fields, we thus determined  the independent, 
tunable  quantities  characterizing Lifshitz black holes.

We were able to formulate a precise analytical understanding as 
to how asymptotically Lifshitz configurations reduce to adS ones, in the 
limit $z\to 1$, which could  be helpful  in the future for
comparing  predictions of field theory duals of adS and Lifshitz theories. 
Then, we studied numerically how the asymptotics match  
black hole horizons at finite values of the radial coordinates, 
computing two-parameter families of black holes, for 
both adS and Lifshitz asymptotics.   

One common feature displayed by the Lifshitz black holes is the presence
of a non-monotonicity in the time and radial metric potentials,
$F$ and $D$, particularly well illustrated in figure \ref{fig:NumLifz}.
This peaking of the Newtonian potential becomes extremely marked
at lower $\varphi_0$ and ${\hat p}_0$, with $F$ peaking at over 
$50$ or higher. 
Although it might appear as if a large peak in $F$ could provide
a barrier to approaching the event horizon, in fact this is
outweighed by the strong Lifshitz $r^{2z}$ warp factor, and
particles are always pulled into the black hole.

As well as studying the field profiles for the black hole 
solutions, and how they vary with $B$-charge, scalar charge 
and dynamical exponent $z$, we initiated a study of the 
thermodynamics of our black holes.
We find the $B$-charge to play an analagous role to the charge 
in the Reissner-Nordstr\"om black hole, with temperature 
decreasing as the charge increases, however, the zero charge limit 
is more subtle for the Lifshitz black holes.  Meanwhile, 
adS/Lifshitz black holes exist only for a finite range of 
scalar charge, due to the Liouville type scalar potential, 
and in general for a given temperature there are two possible 
scalar charges, perhaps surprisingly the smaller one being entropically 
preferred.  The interplay between the scalar and $B$-charge is 
also interesting, in particular a high scalar charge reduces 
the amount of possible $B$-charge.

Since we are interested in holographic condensed matter applications, 
we have focussed on planar black hole geometries,  corresponding to 
a finite temperature boundary theory in flat 2+1 dimensions.  
In simpler models, with pure geometry planar black holes, the 
scaling symmetry renders black holes with different horizon sizes, and 
thus temperatures, physically equivalent.
However, in our supergravity construction the necessary, 
additional scales of $B$-charge and scalar charge introduce 
the possibility of phase transitions.  It would be interesting 
to investigate this possibility further, as well as their 
interpretation in the dual field theory\footnote{Phase 
transitions in a phenomenological model that strongly 
resembles our stringy supergravity setup have been studied very 
recently in \cite{hoker}.}. 
Moreover, the extra fields present in the supergravity theory, 
which we have chosen not to activate, represent other possibilities 
for interesting black hole charges and phase transitions.

Another aspect that deserves further thought would be the 
brane interpretation of our type IIA configuration.  In 
the end, we hope that having explicit string theory embeddings 
of Lifshitz black hole geometries will help to develop the 
holographic description of interesting anisotropic condensed matter systems.

\acknowledgments

We would like to thank Harry Braviner, Anton Faedo and 
Simon Ross for discussions. LB was supported by an STFC studentship. 
RG is supported in part by STFC (Consolidated Grant ST/J000426/1),
in part by the Wolfson Foundation and Royal Society, and in part
by Perimeter Institute for Theoretical Physics. 
Research at Perimeter Institute is supported by the Government of
Canada through Industry Canada and by the Province of Ontario through the
Ministry of Economic Development and Innovation.
SLP is funded by the Deutsche Forschungsgemeinschaft (DFG) 
inside the ``Graduiertenkolleg GRK 1463''.
GT is supported by an STFC Advanced Fellowship ST/H005498/1. 
We would also like to thank the Aspen Center for Physics for 
hospitality via the Working Group Program.

\appendix
\section{The autonomous dynamical system}
\label{appds}

In this appendix we reformulate  the  system of supergravity equations
of motion as an autonomous dynamical system. This requires the use of a 
different gauge to the numerical work, however the translation between the
gauges is straightforward. Choosing $d = -\phi/\sqrt{2} -\ln Lg$, in
the general planar metric (\ref{genplang}), ensures that only negative
exponents of $\phi$ appear in the equations of motion. Then, defining
\bea
&&X_1 = \sqrt{\frac{m}{g}} e^{-\sqrt{2}\phi} \;\;,\;\;\;
X_2 = \frac{q}{g} e^{-\sqrt{2}\phi-2h} \;\;, \;\;\;
X_3 = \sqrt{2}\phi' \;\;,\;\;\; X_4 = 2f' \;\;,\;\;\;
\nonumber \\ 
&&
X_5 = 2h' + \sqrt{2}\phi'\;\;,\;\;\; 
X_6 = 2c'\;\;,\;\;\;
X_7 = e^{-2c+\sqrt{2}\phi} P \;\;,\;\;\;
X_8 = e^{-2c+\sqrt{2}\phi} P'\;\;\;\;
\eea
yields the dynamical system
\bea
X_1' &=& - X_1 X_3 \qquad \qquad
X_2' = - X_2 X_5 \label{DSX12} \\
X_3' &=& - X_2^2 + X_7^2 \left ( X_2^2 - \frac{X_1^4}{4} \right )
- \frac{1}{4} + X_1^2 - \frac34 X_1^4 \nonumber \\
&& - X_3 \left ( \frac{X_4 -X_3}{2} + X_5 + X_6\right ) 
+ \frac{X_8^2}{2} \label{DSX3} \\
X_4' &=& X_2^2 + X_7^2 \left ( 3X_2^2 + \frac{X_1^4}{4} \right )
+ \frac{1}{4} + X_1^2 - \frac{X_1^4}{4} \nonumber \\
&& - X_4 \left ( \frac{X_4 -X_3}{2} + X_5 + X_6 \right ) 
+ \frac{X_8^2}{2} \label{DSX4} \\
X_5' &=& (X_7^2-1) \left ( 2X_2^2 + \frac{X_1^4}{2} \right )
- \frac{1}{2} + \frac{X_8^2}{2} +X_4X_6 + \frac{X_6^2}{2} \nonumber \\
&& + (X_5-2X_3) \left (\frac{X_4}{2} + X_6 \right )
-\frac{X_3^2 +X_3X_5 + X_5^2}{2} \label{DSX5} \\
X_6' &=& X_2^2 - X_7^2 \left ( X_2^2 + \frac{3 X_1^4}{4} \right )
+ \frac{1}{4} + X_1^2 - \frac{X_1^4}{4} \nonumber \\
&& - X_6 \left ( \frac{X_4 -X_3}{2} + X_5 + X_6 \right ) 
- \frac{X_8^2}{2} \label{DSX6} \\
X_7' &=& X_8 - X_6 X_7 + X_3 X_7 \label{DSX7} \\
X_8' &=& -X_8( \frac12 X_4 + X_5 + \frac{X_3}{2})
+ X_7 \left [ X_1^4 + 4 X_2^2 \right ]
\label{DSX8} 
\eea
in which the solution lies in an invariant 7D submanifold described by
the constraint
\bea
&& 2 X_4 (X_6 + X_5 - X_3) + 4 (X_5-X_3) X_6 + X_6^2 + X_5^2 - 2X_5X_3
-X_3^2 -X_8^2 \nonumber \\
&=& 4\frac{K}{qg} X_2 -4 X_2^2 - X_7^2 (4 X_2^2
+ X_1^4) + ( 1 + 4 X_1^2 - X_1^4 ) \,.\label{invsub}
\eea

For a critical point, (\ref{DSX12}) implies $X_3 = X_5 = 0$, 
and (\ref{DSX7},\ref{DSX8}) give:
\bea
X_8  &=& X_6 X_7 \\
X_4 X_8 &=& 2 X_7 \left [ X_1^4 + 4 X_2^2 \right ]
\eea
These are solved by either $X_7 = X_8 = 0$, or
\be
X_4 X_6 = 2 X_1^4 + 8 X_2^2 \,.
\label{crit78}
\ee
We can distinguish two cases:

\smallskip

\noindent $\bullet $ {\bf Case 1: adS}

\smallskip

\noindent
In this case
 $X_7=X_8=0$. Solving (\ref{DSX3}-\ref{DSX6}) gives 
\bea
X_1^2 &=& 1 - \sqrt{1-\frac32 X_6^2} \\
X_2^2 &=& \frac98 X_6^2 - \frac34 + \frac12 \sqrt{1-\frac32 X_6^2}\\
X_4 &=& X_6
\eea
with the constraint
\be
\frac{2X_2}{qg} = 3(1-X_6^2) - 2\sqrt{1-\frac32 X_6^2}
\label{adsc}
\ee
selecting two values for $X_6$ for each charge. In terms of
the supergravity parameters, ${\hat g}= 2/X_6$, ${\hat m} = 
2x_1^2/X_6$. To get the solution in the original area gauge, 
note that $r = e^c =e^{X_6\rho/2}$. These critical points form a curve
of adS solutions in the phase space, with the curve intersecting the
invariant submanifold at two points in general.

To analyze the nature of the critical points, one takes small perturbations,
and finds the eigenvalues and eigenvectors of the perturbation operator
matrix given by $\delta X_\alpha' = M_{\alpha\beta} X_\beta$, 
i.e.\ $M v^{(i)} = \lambda_i v^{(i)}$. There will always be one zero
eigenvalue to this matrix, corresponding to moving along the adS solution
curve. The remaining 7th order polynomial can be factorized, yielding
the eigenvalues plotted in figure \ref{fig:Adseval}, although note that
to obtain figure \ref{fig:Adseval}, we have transformed our 
coordinates to the area gauge
to get the $\Delta_i$ to correspond to the $r$ fall-off exponents: $\Delta_i
= 2\lambda_i/X_6 = {\hat g}\lambda_i$.

\smallskip

\noindent $\bullet $ {\bf Case 2: Lifshitz}

\smallskip

\noindent
Here $X_8 = X_6X_7$, and (\ref{crit78}) holds, and solving the
remaining equations gives:
\bea
X_4 &=& X_6 (1+X_7^2) = \sqrt{\frac{2(1+X_7^2)}{(5+X_7^2)}}\\
X_1^2 &=& \frac{(5+X_7^2) \mp \sqrt{2(5+X_7^2)}}{(1+X_7^2)(5+X_7^2)}\\
X_2^2 &=& \frac{(X_7^2 + 3)(X_7^2-2) \pm 2 \sqrt{2(5+X_7^2)}}
{4(5+X_7^2)(1+X_7^2)^2} \,.\label{X2lif}
\eea
For $X_7^2< \sqrt{2} - (1+\sqrt{17+4\sqrt{2}})/2 \simeq3.3$,
in (\ref{X2lif}) only the upper branch choice gives a real solution for $X_2$.
Finally, the constraint determines the charge:
\be
qg = \frac{\sqrt{(X_7^2+5) \left ((X_7^2+3)(X_7^2-2) \pm 2\sqrt{2(X_7^2+5)}
\right ) }} {3(X_7^2+3) \mp 2\sqrt{2(X_7^2+5)}} \,.
\ee
Clearly $X_7^2 = z-1$, and this is equivalent to the exact solution
(\ref{eq:lifsolns}) in the area gauge.

The analysis of the perturbations around the Lifshitz critical points,
although conceptually identical to adS, is algebraically more involved.
The eigenvalues pair around $-(z+2)/2$ (in area gauge) with the variance
given by the square root of the solution of a cubic equation. While
this can be written in closed form, it is a rather long and
unilluminating expression.
Figure \ref{fig:liffp} shows a plot of the eigenvalues
renormalized for a $r$ fall-off: $\Delta_i = \sqrt{2z(4+z)} \lambda_i$.
All of the eigenvectors of the perturbation operator have 
nonzero components in the vector, scalar and geometry directions.

\section{Exact Lifshitz solutions}
\label{appel}

So far in the literature it has been possible to obtain exact 
Lifshitz black hole (LiBH) solutions only in some phenomenological 
models, where the matter content is engineered to support the 
desired geometry.  In this appendix we give a brief account of 
these approaches, and extend them to dilatonic models, which may 
be more easily embedded into supergravity and string theory.  
We then show that such simple analytical solutions cannot be found 
in the Romans' 6D supergravity that is the main subject of this paper.  

Analytic LiBH solutions  have been found for essentially two
types of 4D Einstein gravity systems (see also \cite{R^2} for 
other possible extensions).  The first ($\Lambda {\mathcal A}A m$) 
contains, besides gravity, a cosmological constant, a massless 
abelian gauge field ${\mathcal F}_2$ and massive abelian gauge 
field $F_2$ with mass $m$ \cite{Pang}.  In 4D, the massive 
gauge field is equivalent to a 
2-form $B_2$ and a massless gauge 
field $F_2$ with non trivial Chern-Simons terms $F_2 \wedge B_2$, 
as studied in \cite{Taylor}.
The second system ($\Lambda {\mathcal A} \phi$) consists of gravity,
a cosmological constant,  a number of massless abelian gauge 
fields ${\mathcal F}^i_2$ and a massless
scalar field $\phi$ with dilaton-like couplings to 
the gauge sector \cite{Taylor,Vandoren}.
An obstruction to straightfowardly embedding  these setups 
into supergravity and string theory is their absence of a 
genuine dilaton field.  For example, many supergravity 
theories, like Romans' 6D supergravity, necessarily contain 
a dilaton field in the supergravity multiplet.
Indeed, it seems unlikely that the $\Lambda {\mathcal A} \phi$ 
system could be embedded in string theory, without generalizing 
the cosmological constant to a genuine dilatonic potential.
It is then easy to check that in the
presence of a non-trivial dilaton potential, the dilaton equation
of motion prevents  the Lifshitz asymptotics. This can be 
observed from the general solutions discussed  in  \cite{BNQTZ}.

The $\Lambda {\mathcal A} A m$ system is more interesting.  
Building on \cite{maldacena, STLIFl}, a similar model, though 
without the massless vector and with two additional dilatonic/radion 
scalars, was obtained in \cite{Amado:2011nd} via a consistent 
massive truncation of Type IIB on an arbitrary Einstein space 
times $S^1$, and used to derive numerical stringy LiBHs.
Here, we generalise the system to a general dilatonic theory, 
with generic dilaton-matter couplings and dilaton potential, 
and search for analytical LiBHs.

\subsection{4D LiBHs with constant dilaton}
 
We consider the four dimensional case, which is 
sufficient to illustrate our strategy.  As well as massless 
and massive abelian gauge fields, we add a dilaton
field  $\phi$ with couplings $\lambda, \sigma$ to the gauge 
fields and a general potential $V(\phi)$.
In this section we
follow closely the discussions in \cite{Pang, Vandoren}, and 
assume their mostly plus metric signature conventions. 
 
We take the general action: 
\be
S = \frac{1}{\kappa^2_4} \int{d^4 x \sqrt{g} 
\left[ R -\frac{1}{2}(\partial\phi)^2 - \frac{e^{-\lambda \phi}}{4} {F}_2^2 -
\frac{e^{-\lambda \phi} m^2}{2} {A}_1^2 -\frac{e^{-\sigma \phi}}{4} 
{\mathcal F}_2^2  
- V(\phi) \right]}
\ee
with corresponding equations of motion:
\bea
&& \partial_\mu \left[ \sqrt{-g}e^{-\lambda\phi} {F}^{\mu\nu}  \right] =
\sqrt{-g}\,  e^{-\lambda\phi} m^2 {A}^{\nu} \\
&& \partial_\mu \left[ \sqrt{-g}e^{-\sigma\phi} {\mathcal F}^{\mu\nu}  
\right] =0 \\
&& \frac{1}{\sqrt{-g}}\partial_\mu \left[ \sqrt{-g} g^{\mu\nu} 
\partial_\nu \phi \right] = \frac{\partial V}{\partial \phi} 
-\frac{\lambda e^{-\lambda\phi}}{4}
{F}^2 -\frac{\lambda\,m^2 e^{-\lambda\phi}}{2}\,{A}^2 
-\frac{\sigma e^{-\sigma\phi}}{4}{\mathcal F}^2 \\
&& R_{\mu\nu} = \frac{1}{2}\partial_\mu\phi\partial_\nu\phi 
+\frac{V}{2}g_{\mu\nu}  + 
\frac{e^{-\lambda\phi}}{4}\left[  2{F}_{\mu}^{\,\lambda}{F}_{\mu\lambda}  
- \frac{g_{\mu\nu}}{2} {F}^2   + 2 m^2 {A}_{\mu} {A}_{\nu}  \right] \nonumber \\
&& \qquad \qquad
+ \frac{ e^{-\sigma\phi}}{4}\left[  2{\mathcal F}_{\mu}^{\,\lambda}
{\mathcal F}_{\mu\lambda}  
- \frac{g_{\mu\nu}}{2} {\mathcal F}^2   \right] \,.
\eea
Consider now the metric Ansatz
\be
ds^2 =  -r^{2z} h(r)\, dt^2  + \frac{dr^2}{r^2h(r)} + r^2 dx^2_i\,,
\ee
sourced by ${\mathcal F}_{rt}$, ${\mathcal F}_{{x_1}{x_2}}$ 
and ${F}_{rt}$, plus a 
constant dilaton field $\phi = const$.
Solving the field equations for the forms gives: 
\bea
&& {\mathcal F}_{rt} = Q_1 r^{z-3}  \,, \qquad  
\qquad {\mathcal F}_{{x_1}{x_2}} = Q_2   \\
&& \partial_r \left[  r^{3-z}\,\partial_r {A}_t  \right]  
= \frac{m^2 {A}_t}{r^{z-1} h(r)}\,.
\label{vectoreq}
\eea
From the Einstein equations  $R^t_t- R_r^r $, we obtain the 
solution for the gauge field: 
\be
{A}_{t} = \pm \frac{2}{m}e^{\lambda\phi/2}\,\sqrt{z-1} \, r^z \, h \,,
\ee
and then using (\ref{vectoreq}) we find the metric function:
\be\label{eqh}
h= \frac{m^2}{2\,z} + \frac{C_1}{(z-2)}\frac{1}{r^2} + \frac{C_2}{r^z}\,,
\ee
for $z\neq2$, and an analogous expression for $z=2$.

Now the dilaton and remaining Einstein equations can be solved provided 
that the following constraints among the parameters are satisfied: 
\bea
&& z=4  \\
&& C_1 = 0  \\
&& -C_2 = \frac{1}{8} e^{-\sigma\phi} (Q_2^2 + Q_1^2)   \qquad  
\Rightarrow \qquad C_2 <0 \\
&&
-\frac{m^2}{4 z} \left[  2(z+2) + z(z-1) \right]  = \frac{V}{2} 
\\
&& 2\lambda (z-1)\,C_2 =  \frac{\sigma}{2} e^{-\sigma\phi} (Q_2^2-Q_1^2) \\
&&  \lambda (z-1)\,m^2\left(\frac{1}{2}+\frac{1}{z}\right) 
= -\frac{\partial V}{\partial\phi}   \,.
\eea
The first four constraints above correspond to those in \cite{Pang} where 
there is no dilaton, and the dilaton adds two more.  In the end 
there are four non-trivial constraints on the four solution 
parameters $C_2$, $Q_1$, $Q_2$ and $\phi$, plus the mass 
parameter $m$ and any gauge couplings that appear in $V(\phi)$.  
So, provided that these constraints can be solved consistently, we 
can avoid tuning the dilaton couplings $\lambda$, $\sigma$.

\subsection{6D Romans' LiBH with constant dilaton?}

Romans' 6D supergravity has a strong resemblance to the dilatonic 
theory we just discussed, or its Chern-Simons equivalent.  
Besides having two extra dimensions, the main difference is that 
the 2-form potential has not only a Chern-Simons term but also a mass term.
Nevertheless, it is straightforward to apply the above strategy to 
search for LiBH solutions with constant dilaton in Romans' supergravity.  
These solutions would be orthogonal to the ones we discuss in 
the main text, since they involve turning on additional fields.  

Indeed, whereas in the main text we activated only one of the 
gauge fields, we now add a non-trivial configuration for the 
second gauge field, taking the Ansatz (\ref{eq:FieldStrength1}) plus:
\be
{\mathcal F}_{rt} = Q_1 r^{z-3}\, \quad {\mathcal F}_{{x_1}{x_2}} = Q_2 \,.
\ee
The expression for the metric function that follows is:
\be
h(r) = \frac{{\rm m}\, Q_2 \,e^{-2\sqrt{2}\phi_0}}{(z-2)\sqrt{z-1}}
\frac{1}{r^2} + \frac{C_1}{r^z} + \frac{L^2}{2z} \left[{\rm m}^2 
e^{-3\sqrt{2}\phi_0} + 4 \frac{q^2}{a^4} e^{-\sqrt{2}\phi_0}\right]\, ,
\ee
for $z\neq2$, and an analogous expression for $z=2$.  
Unfortunately, the constraints coming from the field equations can only 
be solved by a pure Lifshitz configuration, with $Q_1 = 0 = Q_2$ 
and $C_1 = 0$, and the relations (\ref{eq:lifsolns}).
Thus we cannot construct a simple analytic solution via this method.


%

%

\begin{thebibliography}{99}

\bibitem{Malda}
J.~M.~Maldacena,
Adv.\ Theor.\ Math.\ Phys.\  {\bf 2}, 231 (1998)
[Int.\ J.\ Theor.\ Phys.\  {\bf 38}, 1113 (1999)]
[arXiv:hep-th/9711200].

\bibitem{CMP}
S.~A.~Hartnoll,
``Lectures on holographic methods for condensed matter physics,''
Class.\ Quant.\ Grav.\  {\bf 26}, 224002 (2009)
[arXiv:0903.3246 [hep-th]].\\
S.~Sachdev,
``Condensed matter and AdS/CFT,''
arXiv:1002.2947 [hep-th].

\bibitem{SCH}
D.~T.~Son,
Phys.\ Rev.\  D {\bf 78}, 046003 (2008)
[arXiv:0804.3972 [hep-th]].\\
K.~Balasubramanian and J.~McGreevy,
Phys.\ Rev.\ Lett.\  {\bf 101}, 061601 (2008)
[arXiv:0804.4053 [hep-th]].

\bibitem{KLM}
S.~Kachru, X.~Liu and M.~Mulligan,
``Gravity Duals of Lifshitz-like Fixed Points,''
Phys.\ Rev.\  D {\bf 78} (2008) 106005
[arXiv:0808.1725 [hep-th]].

\bibitem{Taylor}
M.~Taylor,
``Non-relativistic holography,''
arXiv:0812.0530 [hep-th].

\bibitem{CM}
K.~Copsey and R.~Mann,
JHEP {\bf 1103}, 039 (2011)
[arXiv:1011.3502 [hep-th]].\\
G.~T.~Horowitz and B.~Way,
``Lifshitz Singularities,''
arXiv:1111.1243 [hep-th].

\bibitem{STLIFe}
S.~A.~Hartnoll, J.~Polchinski, E.~Silverstein and D.~Tong,
``Towards strange metallic holography,''
JHEP {\bf 1004}, 120 (2010)
[arXiv:0912.1061 [hep-th]].
J.~Blaback, U.~H.~Danielsson and T.~Van Riet,
``Lifshitz backgrounds from 10d supergravity,''
JHEP {\bf 1002}, 095 (2010)
[arXiv:1001.4945 [hep-th]].
K.~Balasubramanian and K.~Narayan,
``Lifshitz spacetimes from AdS null and cosmological solutions,''
JHEP {\bf 1008}, 014 (2010)
[arXiv:1005.3291 [hep-th]].

\bibitem{STLIFl}
A.~Donos and J.~P.~Gauntlett,
JHEP {\bf 1012}, 002 (2010)
[arXiv:1008.2062 [hep-th]].
A.~Donos, J.~P.~Gauntlett, N.~Kim and O.~Varela,
JHEP {\bf 1012}, 003 (2010)
[arXiv:1009.3805 [hep-th]].
D.~Cassani and A.~F.~Faedo,
JHEP {\bf 1105}, 013 (2011)
[arXiv:1102.5344 [hep-th]].
  W.~Chemissany and J.~Hartong,
  Class.\ Quant.\ Grav.\  {\bf 28} (2011) 195011
  [arXiv:1105.0612 [hep-th]].



\bibitem{GPTZ}
R.~Gregory, S.~L.~Parameswaran, G.~Tasinato and I.~Zavala,
JHEP {\bf 1012}, 047 (2010)
[arXiv:1009.3445 [hep-th]].

\bibitem{romans6d}
L.~J.~Romans,
Nucl.\ Phys.\  B {\bf 269} (1986) 691.

\bibitem{romans5d}
L.~J.~Romans,
Nucl.\ Phys.\  B {\bf 267}, 433 (1986).

\bibitem{IIAred}
M.~Cvetic, H.~Lu and C.~N.~Pope,
Phys.\ Rev.\ Lett.\  {\bf 83} (1999) 5226
[arXiv:hep-th/9906221].

\bibitem{IIBred}
H.~Lu, C.~N.~Pope and T.~A.~Tran,
Phys.\ Lett.\  B {\bf 475} (2000) 261
[arXiv:hep-th/9909203].
M.~Cvetic, H.~Lu and C.~N.~Pope,
Nucl.\ Phys.\  B {\bf 597} (2001) 172
[arXiv:hep-th/0007109].



\bibitem{Singh}
  H.~Singh,
  Phys.\ Lett.\  B {\bf 682} (2009) 225
  [arXiv:0909.1692 [hep-th]].

\bibitem{HPZ}
  N.~Halmagyi, M.~Petrini and A.~Zaffaroni,
  JHEP {\bf 1108} (2011) 041
  [arXiv:1102.5740 [hep-th]].


\bibitem{Danielsson:2009gi}
U.~H.~Danielsson and L.~Thorlacius,
JHEP {\bf 0903}, 070 (2009)
[arXiv:0812.5088 [hep-th]].


\bibitem{Pang}
D.-W.~Pang,
JHEP {\bf 1001} (2010) 116
[arXiv:0911.2777 [hep-th]].
 
\bibitem{Vandoren}
J.~Tarr\'io and S.~Vandoren,
JHEP {\bf 1109} (2011) 017
[arXiv:1105.6335 [hep-th]].

\bibitem{LiBHpheno}
R.~B.~Mann,
JHEP {\bf 0906} (2009) 075
[arXiv:0905.1136 [hep-th]]\,;\\
G.~Bertoldi, B.~A.~Burrington and A.~Peet,
Phys.\ Rev.\  D {\bf 80}, 126003 (2009)
[arXiv:0905.3183 [hep-th]] \,;\\
G.~Bertoldi, B.~A.~Burrington and A.~W.~Peet,
Phys.\ Rev.\  D {\bf 80} (2009) 126004
[arXiv:0907.4755 [hep-th]\\
E.~J.~Brynjolfsson, U.~H.~Danielsson, L.~Thorlacius and T.~Zingg,
J.\ Phys.\ A  {\bf 43} (2010) 065401
[arXiv:0908.2611 [hep-th]] \,;\\
M.~C.~N.~Cheng, S.~A.~Hartnoll and C.~A.~Keeler,
JHEP {\bf 1003} (2010) 062
[arXiv:0912.2784 [hep-th]];\\
E.~Ayon-Beato, A.~Garbarz, G.~Giribet and M.~Hassaine,
JHEP {\bf 1004} (2010) 030
[arXiv:1001.2361 [hep-th]] \,;\\
G.~Bertoldi, B.~A.~Burrington and A.~W.~Peet,
Phys.\ Rev.\  D {\bf 82}, 106013 (2010)
[arXiv:1007.1464 [hep-th]] \,;\\
G.~Bertoldi, B.~A.~Burrington, A.~W.~Peet and I.~G.~Zadeh,
Phys.\ Rev.\  D {\bf 83}, 126006 (2011)
[arXiv:1101.1980 [hep-th]] \,;\\ 
M.~H.~Dehghani, R.~B.~Mann and R.~Pourhasan,
Phys.\ Rev.\  D {\bf 84} (2011) 046002
[arXiv:1102.0578 [hep-th]].

\bibitem{R^2}
R.~B.~Mann,
JHEP {\bf 0906} (2009) 075
[arXiv:0905.1136 [hep-th]];\\
K.~Balasubramanian and J.~McGreevy,
Phys.\ Rev.\ D {\bf 80} (2009) 104039
[arXiv:0909.0263 [hep-th]];\\
E.~Ayon-Beato, A.~Garbarz, G.~Giribet and M.~Hassaine,
Phys.\ Rev.\ D {\bf 80} (2009) 104029
[arXiv:0909.1347 [hep-th]];\\
R.~-G.~Cai, Y.~Liu and Y.~-W.~Sun,
JHEP {\bf 0910} (2009) 080
[arXiv:0909.2807 [hep-th]];\\
M.~H.~Dehghani and R.~B.~Mann,
JHEP {\bf 1007} (2010) 019
[arXiv:1004.4397 [hep-th]];\\
W.~G.~Brenna, M.~H.~Dehghani and R.~B.~Mann,
Phys.\ Rev.\  D {\bf 84} (2011) 024012
[arXiv:1101.3476 [hep-th]];\\
H.~Maeda and G.~Giribet,
JHEP {\bf 1111} (2011) 015
[arXiv:1105.1331 [gr-qc]].
  J.~Matulich and R.~Troncoso,
  JHEP {\bf 1110} (2011) 118
  [arXiv:1107.5568 [hep-th]].
 
 
\bibitem{Amado:2011nd}
I.~Amado and A.~F.~Faedo,
JHEP {\bf 1107}, 004 (2011)
[arXiv:1105.4862 [hep-th]].

  
\bibitem{MN}
J.~M.~Maldacena and C.~N\'u\~nez,
``Supergravity description of field theories on curved manifolds and 
a no  go theorem,''
Int.\ J.\ Mod.\ Phys.\  A {\bf 16} (2001) 822
[arXiv:hep-th/0007018].

\bibitem{BGR}
H.~Braviner, R.~Gregory and S.~F.~Ross,
Class.\ Quant.\ Grav.\  {\bf 28}, 225028 (2011)
[arXiv:1108.3067 [hep-th]].


\bibitem{BF}
P.~Breitenlohner and D.~Z.~Freedman,
Phys.\ Lett.\  B {\bf 115}, 197 (1982).\\
P.~Breitenlohner and D.~Z.~Freedman,
Annals Phys.\  {\bf 144}, 249 (1982).


\bibitem{hoker}
E.~D'Hoker and P.~Kraus,
``Charge Expulsion from Black Brane Horizons, and Holographic 
Quantum Criticality in the Plane,''
arXiv:1202.2085 [hep-th].

\bibitem{maldacena}
J.~Maldacena, D.~Martelli and Y.~Tachikawa,
JHEP {\bf 0810} (2008) 072
[arXiv:0807.1100 [hep-th]].

\bibitem{BNQTZ}
C.~P.~Burgess, C.~N\'u\~nez, F.~Quevedo, G.~Tasinato and I.~Zavala,
JHEP {\bf 0308} (2003) 056
[arXiv:hep-th/0305211].

\end{thebibliography}
\end{document}